\documentclass[twocolumn,iop]{emulateapj}
\usepackage{apjfonts}
\usepackage{color}

\usepackage{amsmath,graphicx,longtable,hyperref,lineno}
\usepackage{natbib,threeparttable,rotating}
\usepackage{ulem}

\newcommand{\etal}{et al.}
\newcommand{\hbeta}{H{$\beta$}}

\newcommand{\CIV}{C\,{\sevenrm IV}}

\def\FeII{Fe\,{\sc ii}}
\def\MgII{Mg\,{\sc ii}}

\def \OIII {[O\,{\sc iii}]}

   \font\sevenrm=cmr7 scaled 1000
\newcommand{\comments}[1]{}

\def\kms{{\rm km\,s^{-1}}}

\begin{document}

\title{Extreme variability quasars from the Sloan Digital Sky Survey and the Dark Energy Survey}

\def\andname{}
\author{
N.~Rumbaugh\altaffilmark{1},
Yue~Shen\altaffilmark{2,1,*},
Eric~Morganson\altaffilmark{1},
Xin~Liu\altaffilmark{2},
M.~Banerji\altaffilmark{3,4},
R.~G.~McMahon\altaffilmark{3,4},
F.~B.~Abdalla\altaffilmark{5,6},
A.~Benoit-L{\'e}vy\altaffilmark{7,5,8},
E.~Bertin\altaffilmark{7,8},
D.~Brooks\altaffilmark{5},
E.~Buckley-Geer\altaffilmark{9},
D.~Capozzi\altaffilmark{10},
A. Carnero Rosell\altaffilmark{11,12},
M.~Carrasco~Kind\altaffilmark{2,1},
J.~Carretero\altaffilmark{13},
C.~E.~Cunha\altaffilmark{14},
C.~B.~D'Andrea\altaffilmark{15},
L.~N.~da Costa\altaffilmark{11,12},
D.~L.~DePoy\altaffilmark{16},
S.~Desai\altaffilmark{17},
P.~Doel\altaffilmark{5},
J.~Frieman\altaffilmark{9,18},
J.~Garc\'ia-Bellido\altaffilmark{19},
D.~Gruen\altaffilmark{14,20},
R.~A.~Gruendl\altaffilmark{2,1},
J.~Gschwend\altaffilmark{11,12},
G.~Gutierrez\altaffilmark{9},
K.~Honscheid\altaffilmark{21,22},
D.~J.~James\altaffilmark{23,24},
K.~Kuehn\altaffilmark{25},
S.~Kuhlmann\altaffilmark{26},
N.~Kuropatkin\altaffilmark{9},
M.~Lima\altaffilmark{27,11},
M.~A.~G.~Maia\altaffilmark{11,12},
J.~L.~Marshall\altaffilmark{16},
P.~Martini\altaffilmark{21,28},
F.~Menanteau\altaffilmark{2,1},
A.~A.~Plazas\altaffilmark{29},
K.~Reil\altaffilmark{20},
A.~Roodman\altaffilmark{14,20},
E.~Sanchez\altaffilmark{30},
V.~Scarpine\altaffilmark{9},
R.~Schindler\altaffilmark{20},
M.~Schubnell\altaffilmark{31},
E.~Sheldon\altaffilmark{32},
M.~Smith\altaffilmark{33},
M.~Soares-Santos\altaffilmark{9},
F.~Sobreira\altaffilmark{34,11},
E.~Suchyta\altaffilmark{35},
M.~E.~C.~Swanson\altaffilmark{1},
A.~R.~Walker\altaffilmark{24},
W.~Wester\altaffilmark{9}
\\ \vspace{0.2cm} (DES Collaboration) \\
}

\affil{$^{1}$ National Center for Supercomputing Applications, 1205 West Clark St., Urbana, IL 61801, USA}
\affil{$^{2}$ Department of Astronomy, University of Illinois, 1002 W. Green Street, Urbana, IL 61801, USA}
\affil{$^{3}$ Institute of Astronomy, University of Cambridge, Madingley Road, Cambridge CB3 0HA, UK}
\affil{$^{4}$ Kavli Institute for Cosmology, University of Cambridge, Madingley Road, Cambridge CB3 0HA, UK}
\affil{$^{5}$ Department of Physics \& Astronomy, University College London, Gower Street, London, WC1E 6BT, UK}
\affil{$^{6}$ Department of Physics and Electronics, Rhodes University, PO Box 94, Grahamstown, 6140, South Africa}
\affil{$^{7}$ CNRS, UMR 7095, Institut d'Astrophysique de Paris, F-75014, Paris, France}
\affil{$^{8}$ Sorbonne Universit\'es, UPMC Univ Paris 06, UMR 7095, Institut d'Astrophysique de Paris, F-75014, Paris, France}
\affil{$^{9}$ Fermi National Accelerator Laboratory, P. O. Box 500, Batavia, IL 60510, USA}
\affil{$^{10}$ Institute of Cosmology \& Gravitation, University of Portsmouth, Portsmouth, PO1 3FX, UK}
\affil{$^{11}$ Laborat\'orio Interinstitucional de e-Astronomia - LIneA, Rua Gal. Jos\'e Cristino 77, Rio de Janeiro, RJ - 20921-400, Brazil}
\affil{$^{12}$ Observat\'orio Nacional, Rua Gal. Jos\'e Cristino 77, Rio de Janeiro, RJ - 20921-400, Brazil}
\affil{$^{13}$ Institut de F\'{\i}sica d'Altes Energies (IFAE), The Barcelona Institute of Science and Technology, Campus UAB, 08193 Bellaterra (Barcelona) Spain}
\affil{$^{14}$ Kavli Institute for Particle Astrophysics \& Cosmology, P. O. Box 2450, Stanford University, Stanford, CA 94305, USA}
\affil{$^{15}$ Department of Physics and Astronomy, University of Pennsylvania, Philadelphia, PA 19104, USA}
\affil{$^{16}$ George P. and Cynthia Woods Mitchell Institute for Fundamental Physics and Astronomy, and Department of Physics and Astronomy, Texas A\&M University, College Station, TX 77843,  USA}
\affil{$^{17}$ Department of Physics, IIT Hyderabad, Kandi, Telangana 502285, India}
\affil{$^{18}$ Kavli Institute for Cosmological Physics, University of Chicago, Chicago, IL 60637, USA}
\affil{$^{19}$ Instituto de Fisica Teorica UAM/CSIC, Universidad Autonoma de Madrid, 28049 Madrid, Spain}
\affil{$^{20}$ SLAC National Accelerator Laboratory, Menlo Park, CA 94025, USA}
\affil{$^{21}$ Center for Cosmology and Astro-Particle Physics, The Ohio State University, Columbus, OH 43210, USA}
\affil{$^{22}$ Department of Physics, The Ohio State University, Columbus, OH 43210, USA}
\affil{$^{23}$ Astronomy Department, University of Washington, Box 351580, Seattle, WA 98195, USA}
\affil{$^{24}$ Cerro Tololo Inter-American Observatory, National Optical Astronomy Observatory, Casilla 603, La Serena, Chile}
\affil{$^{25}$ Australian Astronomical Observatory, North Ryde, NSW 2113, Australia}
\affil{$^{26}$ Argonne National Laboratory, 9700 South Cass Avenue, Lemont, IL 60439, USA}
\affil{$^{27}$ Departamento de F\'isica Matem\'atica, Instituto de F\'isica, Universidade de S\~ao Paulo, CP 66318, S\~ao Paulo, SP, 05314-970, Brazil}
\affil{$^{28}$ Department of Astronomy, The Ohio State University, Columbus, OH 43210, USA}
\affil{$^{29}$ Jet Propulsion Laboratory, California Institute of Technology, 4800 Oak Grove Dr., Pasadena, CA 91109, USA}
\affil{$^{30}$ Centro de Investigaciones Energ\'eticas, Medioambientales y Tecnol\'ogicas (CIEMAT), Madrid, Spain}
\affil{$^{31}$ Department of Physics, University of Michigan, Ann Arbor, MI 48109, USA}
\affil{$^{32}$ Brookhaven National Laboratory, Bldg 510, Upton, NY 11973, USA}
\affil{$^{33}$ School of Physics and Astronomy, University of Southampton,  Southampton, SO17 1BJ, UK}
\affil{$^{34}$ Instituto de F\'isica Gleb Wataghin, Universidade Estadual de Campinas, 13083-859, Campinas, SP, Brazil}
\affil{$^{35}$ Computer Science and Mathematics Division, Oak Ridge National Laboratory, Oak Ridge, TN 37831}
\affil{* Alfred P. Sloan Research Fellow}

\shorttitle{Extreme Variability Quasars}
\shortauthors{Rumbaugh \etal}

\begin{abstract}
We perform a systematic search for long-term extreme variability quasars (EVQs) in the overlapping Sloan Digital Sky Survey (SDSS) and 3-Year Dark Energy Survey (DES) imaging, which provide light curves spanning more than 15 years. We identified $\sim 1000$ EVQs with a maximum $g$ band magnitude change of more than 1 mag over this period, about $10\%$ of all quasars searched. The EVQs have $L_{\rm bol}\sim10^{45}-10^{47}$\ erg s$^{-1}$ and\ $L/L_{\rm Edd} \sim 0.01-1$. Accounting for selection effects, we estimate an intrinsic EVQ fraction of $\sim 30-50\%$ among all $g\lesssim 22$ quasars over a baseline of $\sim 15$ years. These EVQs are good candidates for so-called ``changing-look quasars'', where a spectral transition between the two types of quasars (broad-line and narrow-line) is observed between the dim and bright states. We performed detailed multi-wavelength, spectral and variability analyses for the EVQs and compared to their parent quasar sample. We found that EVQs are distinct from a control sample of quasars matched in redshift and optical luminosity: (1) their UV broad emission lines have larger equivalent widths; (2) their Eddington ratios are systematically lower; and (3) they are more variable on all timescales. The intrinsic difference in quasar properties for EVQs suggest that internal processes associated with accretion are the main driver for the observed extreme long-term variability. However, despite their different properties, EVQs seem to be in the tail of a continuous distribution of quasar properties, rather than standing out as a distinct population. We speculate that EVQs are normal quasars accreting at relatively low accretion rates, where the accretion flow is more likely to experience instabilities that drive the factor of few changes in flux on multi-year timescales.  
\keywords{
black hole physics -- galaxies: active -- line: profiles -- quasars: general -- surveys
}
\end{abstract}

\section{Introduction}\label{sec:intro}

In the canonical unification picture of Active Galactic Nuclei (AGN) \citep[e.g.,][]{Antonucci_1993,Urry_Padovani_1995}, broad-line (Type 1) and narrow-line (Type 2) objects are the same system of accreting supermassive black holes viewed at different orientations. When the system is viewed nearly edge on, the emission from the accretion disk and the broad-line region (BLR) is blocked by an optically thick dust torus, and the system will appear as a Type 2 AGN \citep[for a recent review on the AGN dust torus, see, e.g.,][]{Netzer_2015}. 

The continuum emission from the accretion disk, which powers the broad-line emission, can vary on timescales of days to years due to fluctuations in the accretion disk \citep[e.g.,][]{Peterson_2001,Armitage_Reynolds_2003,Kelly_etal_2009,Dexter_Agol_2011,Ruan_etal_2014}. The typical amplitude of the optical continuum variability is $\sim 0.2$ mag \citep[e.g.,][]{Vandenberk_etal_2004,Sesar_etal_2007}, although it depends on the timescale, wavelength, and AGN properties such as luminosity and Eddington ratio \citep[e.g.,][]{Schmidt_etal_2010,Ai_etal_2010,Butler_Bloom_2011,MacLeod_etal_2012}. However, large flux variation of a magnitude or more in the continuum is uncommon to observe in AGN and requires a dramatic change in the accretion rate or in the obscuration structure. 

Early repeated observations of low-luminosity AGN have revealed several examples where the continuum and broad-line flux varied by a large factor, {characteristic of a type transition (e.g., Type 1 to Type 2 and vice versa) \citep[e.g.,][]{Goodrich_1995}}. One such example is NGC 4151, where the broad emission lines had disappeared and later reappeared over the course of several decades \citep[e.g.,][]{Osterbrock_1977, Antonucci_Cohen_1983, Lyutyi_etal_1984, Penston_Perez_1984, Shapovalova_etal_2010}. These objects, dubbed ``changing-look'' AGN, have been discovered in greater numbers, and at higher redshifts and higher luminosities (e.g., the quasar regime)\footnote{In this paper we collectively refer to these objects as ``changing-look quasars'' for simplicity. } over the past few years, mostly as a result from large-area optical imaging and spectroscopic surveys \citep[e.g.,][]{Shappee_etal_2014,Denney_etal_2014,LaMassa_etal_2015,MacLeod_etal_2016,Parker_etal_2016,Runco_etal_2016,Runnoe_etal_2016,Ruan_etal_2016,Gezari_etal_2017}. The sample size of these changing-look objects, however, remains small (i.e., only $\sim$ a dozen objects known so far).

The two common interpretations of changing-look quasars (CLQs) are: (1) changes in the accretion rate; (2) changes in the obscuration. Recent studies disfavor the obscuration interpretation in most CLQs discovered to date\footnote{Some AGN show sudden, large changes in their X-ray flux, accompanied by significant changes in the X-ray absorption column density. Such events can be reasonably explained by the occultation of fast-moving gas clouds within the BLR that absorb the inner X-ray emission along the light-of-sight \citep[e.g.,][]{Risaliti_etal_2009}.  } \citep[e.g.,][]{Denney_etal_2014,LaMassa_etal_2015,Husemann_etal_2016,Koay_etal_2016,MacLeod_etal_2016,Ruan_etal_2016,Runnoe_etal_2016,Gezari_etal_2017}. For example, the dust reddening model is unable to simultaneously fit the reduction in both the continuum and the broad-line flux in the dim state; in addition, the broad-line flux is often preferentially reduced in the low-velocity part, which contradicts the obscuration scenario. Other transient scenarios, such as tidal disruption events \citep[][]{Merloni_etal_2015}  or microlensing, also have difficulties to explain the overall observations in the majority of CLQs, although they may be viable in specific cases. Therefore most of these recent studies concluded that changes in the accretion rate is the dominant mechanism for the CLQ phenomenon. 

One significant challenge to the above interpretation is that the timescales over which a changing-look event occurs (e.g.,  less than a few years in the quasar restframe) are typically much shorter than the timescales over which the accretion rate is expected to change by a large factor. The relevant timescale associated with changes in the accretion rate is the viscous timescale \citep[e.g.,][]{Krolik_1999}, which is typically on the order of $\sim 10^4$ yrs given typical accretion parameters of quasars \citep[e.g., Eqn.\, 1 in][]{MacLeod_etal_2016}. Thus only gradual evolution from Type 1 objects to Type 2 objects as accretion rate diminishes may be possible \citep[e.g.,][]{Elitzur_etal_2014}. While the dynamical timescale of the BLR over which the broad-line flux may vary dramatically is $t_{\rm dyn}\approx R_{\rm BLR}/\Delta V\approx $ a few years, where $R_{\rm BLR}$ is the BLR radius and $\Delta V$ is the velocity dispersion in the BLR estimated from the width of the broad lines, the rapid, large-amplitude changes in the accretion rate still lack a theoretical explanation. One possibility is that certain instabilities are operating in the accretion disk and cause large variations of the accretion rate on multi-year timescales. For example, \citet{Jiang_etal_2016} recently proposed that iron opacity in the accretion disk can strongly impact the structure and stability of accretion flows, and may lead to the observed large flux fluctuations over relatively short timescales. 

Given the peculiarities of CLQs and their implications on the accretion processes in quasars, it is important to assemble a large sample of such objects and study their statistical properties and compare them to normal quasars. Since CLQs are an apparently rare phenomenon that occurs on multi-year timescales, the best way to systematically search for these objects is to utilize large-area imaging surveys combined with spectroscopic follow-ups. 

In this work we perform a systematic search for CLQs combining SDSS data and the Dark Energy Survey (DES) Year-3 imaging data for a large sample of quasars in the overlap footprint of SDSS and DES imaging. CLQs are a subset of ``extreme variability quasars'' (EVQs) since the changing-look event is associated with large flux changes between the dim and bright states. We therefore adopt the term EVQs in what follows, and note that a CLQ is loosely defined as a EVQ with observed type transition in the dim and bright states with spectroscopy.  

The time baseline for the combined SDSS and DES multi-epoch photometry spans more than 15 years, ideal for the search of EVQs over long timescales. The large parent sample size and ample multi-wavelength data and spectroscopic measurements of these quasars will allow us to construct the largest sample of EVQs and study their physical properties. 

The paper is organized as follows. In \S\ref{sec:data} we describe the data. In \S\ref{sec:clq} we present the sample of EVQs and explore their multi-wavelength, spectral, and optical variability properties, and compare to normal quasars. We discuss our findings in the context of understanding these objects in \S\ref{sec:disc} and conclude in \S\ref{sec:con}. Throughout the paper we adopt a flat $\Lambda$CDM cosmology with $\Omega_\Lambda=0.7$ and $H_{0}=70\,\kms\,{\rm Mpc}^{-1}$. All magnitudes used are PSF magnitudes in the AB system.

\section{Sample and Data}\label{sec:data}

To search for EVQs we start from the SDSS DR7 quasar catalog \citep[DR7Q,][]{Schneider_etal_2010}, and identify counterparts in the regions overlapping with DES. The SDSS DR7 quasar catalog contains 105,783 quasars with $0.05\lesssim z\lesssim 5$ and luminosities larger than $M_{i}=-22$. All quasars are spectroscopically confirmed, and have a variety of spectral measurements from \citet{Shen_etal_2011}. Roughly half of the quasars in the parent sample were uniformly selected with the final quasar target selection algorithm described in \citet{Richards_etal_2002a} and were targeted to $i=19.1$ (at $z\lesssim 2.9$) and $i=20.2$ (at $z\gtrsim 2.9$). However, the remaining quasars were selected to fainter limiting magnitudes, in particular in the Stripe 82 region (see below). 

The spectral measurements and photometric magnitudes from the \citet{Shen_etal_2011} catalog that we use in our analysis are single-epoch measurements, and can be treated as a random selection from the multi-year light curves. Therefore by using these single-epoch measurements we are probing the average properties of the sample under consideration. 

\subsection{SDSS ($\sim$1998--2009)}\label{sec:sdss}

The SDSS I-II \citep{SDSS} used a dedicated 2.5-m wide-field telescope \citep{Gunn_etal_2006} with a drift-scan camera with 30 $2048\times 2048$ CCDs \citep{Gunn_etal_1998} to image the sky in five broad optical bands \citep[$u\,g\,r\,i\,z$;][]{Fukugita_etal_1996}. The imaging data are taken on dark photometric nights of good seeing \citep{Hogg_etal_2001}, are calibrated photometrically \citep{Smith_etal_2002, Ivezic_etal_2004, Tucker_etal_2006} and astrometrically \citep{Pier_etal_2003}, and object parameters are measured \citep{Lupton_etal_2001}. Quasar candidates \citep{Richards_etal_2002a} for follow-up spectroscopy are selected from the imaging data using their colors, and are arranged in spectroscopic plates \citep{Blanton_etal_2003} to be observed with a pair of double spectrographs \citep{Smee_etal_2013}. 

Most of the photometric data for SDSS DR7 quasars were taken during SDSS I-II, with additional photometry taken as part of the SDSS III survey \citep{Eisenstein_etal_2011}. All available photometric data from the latest SDSS DR13 \citep{DR13} are used in this study. 

There is nominally one SDSS photometric epoch per object. However, in regions where two imaging scans overlap there will be more than one epochs. In addition, a $\sim 270\,{\rm deg^2}$ area ($-50 < RA < 60^{\circ}$, $-1.25 < DEC < 1.25^{\circ}$) along the Celestial Equator, called ``Stripe 82'' (hereafter S82), was repeatedly imaged during $\sim 1998-2009$, producing about 60 epochs for each object \citep[][]{Annis_etal_2014}. Most of the overlap between the SDSS and DES footprints is in the Stripe 82 region for our quasar sample, providing dense S82 light curves to measure the optical variability of quasars over days to multi-year timescales. 

The spectroscopic data used in this study are exclusively from SDSS I-II, which have a wavelength coverage of $\sim 3800-9200$\,\AA\ and a spectral resolution of $R\sim 2000$.  

\begin{figure}
\centering
    \includegraphics[width=0.45\textwidth]{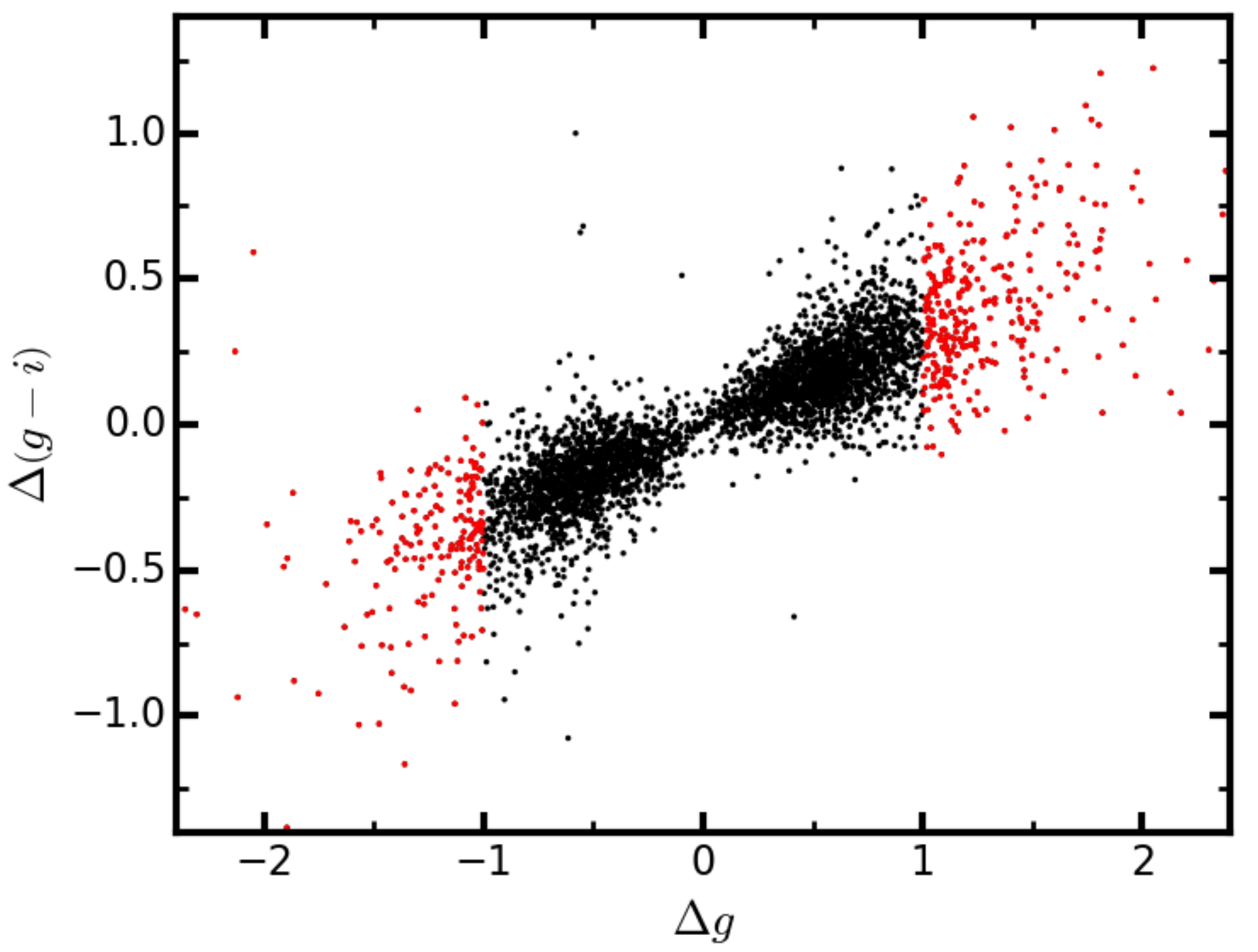}    
    \caption{The $g$ band magnitude change between the first and second extrema is plotted versus the corresponding change in $g-i$, so that a positive $\Delta g$ indicates a decrease in brightness. The red points are the selected EVQs. For quasars at all levels of variability, we see a correlation between changes in magnitude and color. This implies that the $i$-band flux varies in the same direction as the $g$-band flux, but with a reduced amplitude. }
    \label{fig:mag_col_plot}
\end{figure}

\begin{figure}
\centering
    \includegraphics[width=0.28\textheight,angle=-90]{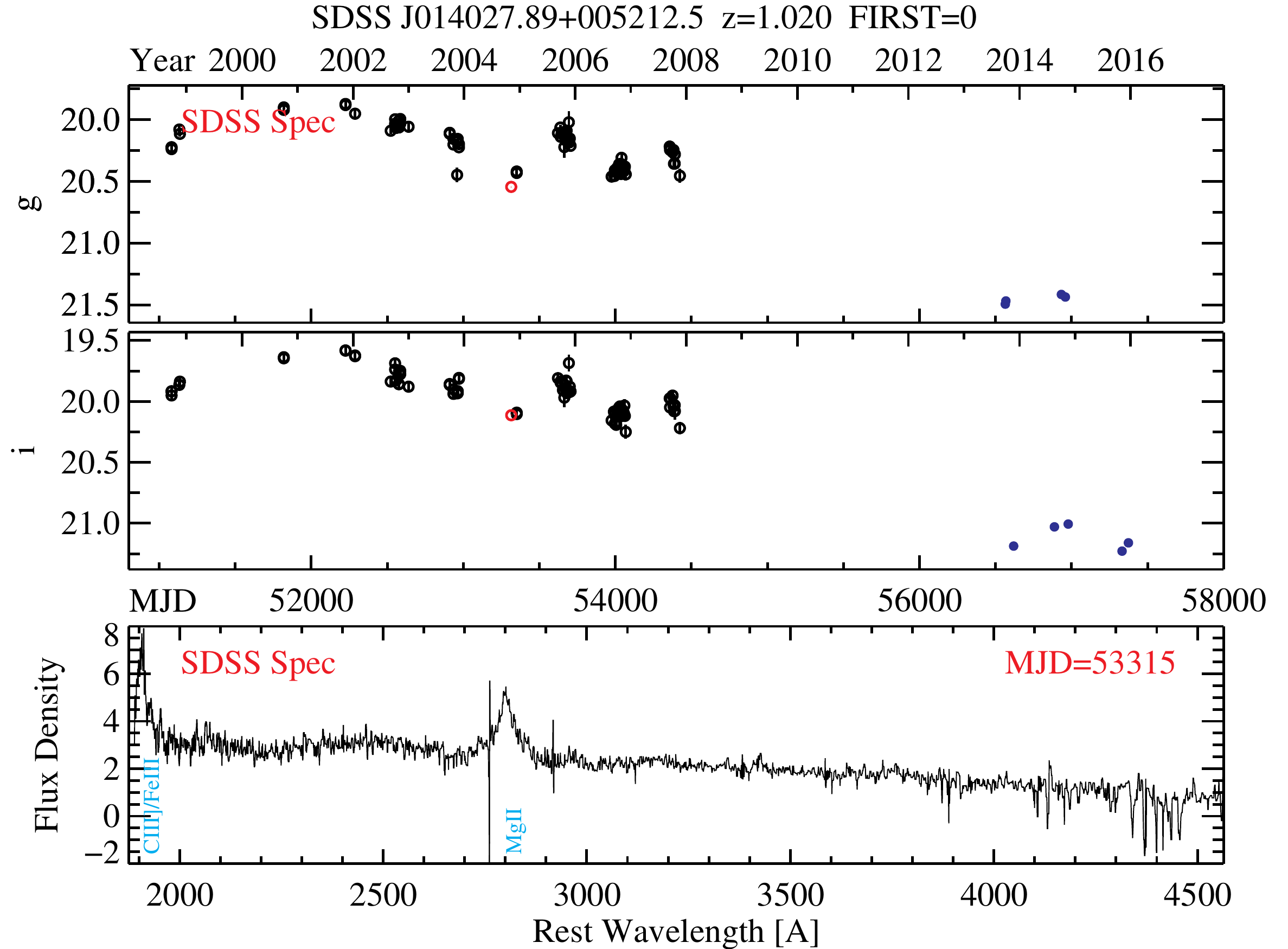}
    \caption{An example EVQ identified from SDSS (MJD$<55000$) and DES (MJD$>56000$) imaging over more than a decade, where the DES epochs are significantly dimmed. The top and middle panels show the $g$ and $i$ light curves, respectively, where the SDSS spectroscopic epoch is indicated by the red circle. The bottom panel shows the SDSS spectrum, where the major broad lines are marked. This object has a flag FIRST$=0$, which means it is undetected in the FIRST radio survey.}
    \label{fig:examp}
\end{figure}

\subsection{DES (Y3A1, $\sim$2013--2016)}\label{sec:des}

The Dark Energy Survey is a wide-area 5000 deg$^2$ {\it grizY} survey of the southern sky \citep{Flaugher_2005,Frieman_etal_2013}. Its primary goal is to uncover the nature of dark energy, using four main cosmological probes: baryon acoustic oscillations, galaxy clusters, weak gravitational lensing, and Type Ia supernovae. The survey is conducted using the Dark Energy Camera \citep[DECam;][]{Flaugher_etal_2015}, a 570 megapixel imager on the 4m Blanco telescope at the Cerro Tololo Inter-American Observatory. DES is deeper than other surveys of similar area, such as SDSS, with typical coadded 5$\sigma$ point source depths of $g=24.7$, $r=24.5$, $i=24.0$, $z=23.3$, and $Y=21.9$ in the first 3 seasons.


{The first season of data collection began in August of 2013, and the third season concluded in February 2016 \citep{Diehl_etal_2016}. For this work, we use the Y3A1 dataset, which includes these first three seasons of observation, as well as some Science Verification data with sufficient image quality. The median number of epochs for our sample is 4 in $g$, 3 in $r$, 3 in $i$, 3 in $z$, and 4 in $Y$. The single-epoch depth is sufficient to detect essentially all SDSS quasars even if they were significantly dimmed. }

\section{Extreme Variability Quasar Selection}\label{sec:clq}

Of the 105,783 quasars in the DR7Q catalog, 8640 were successfully matched to sources in the DES Y3A1 dataset. The statistics of this sample are succinctly summarized in Table \ref{tab:sample}. SDSS quasars are relatively isolated systems with few blending problems with nearby objects. A moderately large matching radius of 2\arcsec\ was used between the SDSS and DES positions, and the nearest match was taken as the same object. The distribution of angular separations of the matches is consistent with that expected from the astrometric uncertainties of SDSS and DES, and these angular separations are typically much less than 1\arcsec. There were 12 matches with angular separations greater than 0.5\arcsec. We checked these objects and found 7 were mismatches in close pairs of objects; we manually corrected these matches. The other 5 objects were due to astrometric errors but are the correct matches. Of these 8640 objects, 7481 are in Stripe 82, where SDSS and DES largely overlap. We found that essentially all SDSS quasars within the DES footprint are detected by DES except for a few objects with processing issues in DES. 

To select EVQs, we use the criteria in \citet{MacLeod_etal_2016} on the combined $g$-band light curves from SDSS$+$DES, and select objects with a magnitude change of $|\Delta g|>1$ mag between any two epochs in the combined light curve. The slight difference in the photometric systems of SDSS and DES can be safely ignored for our purposes. We also require photometric uncertainties $\sigma_g<0.15$ mag to ensure robust measures of flux changes. Before we make the selection, we reject light curve epochs that are apparent outliers. {An epoch was flagged as an outlier when it was at least 0.5 magnitudes away from the running median of all epochs within a window of $\pm$100 days.}

Since the EVQ selection relies on extremes in the light curve, we examined the photometric data to rule out processing issues or potential contamination from nearby objects. The SDSS Stripe 82 light curves were already vetted by \citet{MacLeod_etal_2010}. For DES photometry, we compared the photometric error distribution of the parent quasar sample with that of a comparison star sample with the same magnitude distribution, and found nearly identical distributions. We checked the pipeline processing flags of the EVQs and only found 11 objects whose extremum DES epoch has one of the pipeline extraction flags marked. We inspected these cases individually and concluded the DES photometry and errors are still reliable for these objects. Although we only used $g$ band in the selection of EVQs, we also looked at the $i$ band light curves as an additional check to validate the large flux variability of selected EVQs. We found that changes in $g$ were correlated with changes in $i$, as shown in Fig. \ref{fig:mag_col_plot}, indicating the large flux change in $g$ band is not due to processing issues. Flux changes in $g$ and $r$ were similarly correlated. Finally, for all selected EVQs (including extreme cases with the largest magnitude changes or longest variation baselines), we further visually inspected the combined SDSS$+$DES light curves as well as the image stamps from DES to make sure there are no obvious artifacts in the data that may cause spurious large flux changes. 

Generally speaking, EVQs are not necessarily CLQs, which would require spectroscopic confirmation. However, \citet{MacLeod_etal_2016} showed that $>15\%$ of these EVQs display changing-look features in their broad-line emission on multi-year timescales. If the majority of these EVQs are caused by variations in the accretion rate (and hence the continuum flux), the canonical unification model predicts that the broad-line flux will follow the changes in the continuum flux due to photoionization. 

Fig.\ \ref{fig:examp} shows one example EVQ. The full list of EVQs and their basic properties are provided in Table \ref{tab:evq_sample}.

\subsection{Basic statistics}

\begin{table}
\caption{Sample Statistics}\label{tab:sample}
\centering
\scalebox{1.0}{
\begin{tabular}{cccccc}
\hline\hline
& $N_{\rm DR7Q}$ &  $N_{\rm SDSS+DES}$ & $N_{|\Delta g|>1}$ & $N_{|\Delta g|>1.5}$  & $N_{|\Delta g|>2}$ \\
& (1) & (2)  & (3) & (4) & (5) \\
\hline
all    & 105,783 & 8640  & 977 & 166 & 37 \\
FIRST  &  9399 & 558 & 93 & 25 & 9 \\
S82 &  9258     & 7481  & 898   & 146 & 33     \\  
S82 (FIRST) &  482     & 457  & 81   & 20 & 8     \\    
\hline
\hline\\
\end{tabular}
}
\begin{tablenotes}
      \small
      \item NOTE. --- (1) Total number of SDSS-DR7 quasars; (2) number of DES matches to the SDSS-DR7 quasar catalog; (3)--(5) numbers of selected extreme variability quasars from the combined SDSS and DES light curves with different variability thresholds. The third row shows the corresponding numbers of quasars within the SDSS Stripe 82 region. The second and fourth rows are the same as the first and third rows, respectively, but only for FIRST-detected quasars.
\end{tablenotes}
\end{table}

\begin{figure}
\centering
    \includegraphics[height=0.48\textwidth,angle=-90]{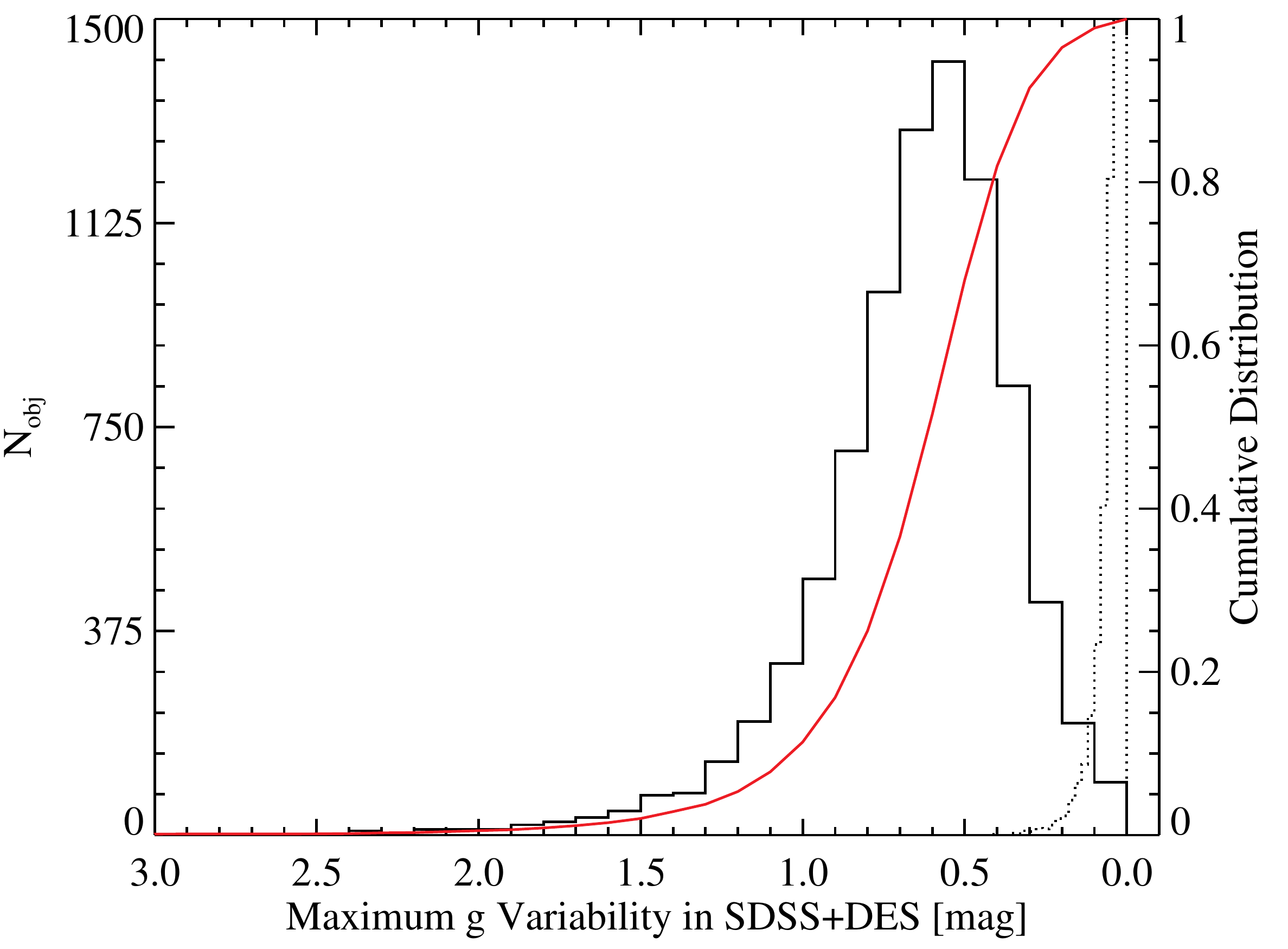}
    \caption{Distribution of the maximum $g$ variability for the SDSS$+$DES matched quasar sample (black solid line). The red line shows the cumulative distribution. The dotted line shows the expected distribution of zero variability convolved with photometric errors, demonstrating that the observed variability is intrinsic. About 10\% of all quasars show maximum $g$ band variability greater than 1 mag from $\sim 16$ years of SDSS and DES imaging. }
    \label{fig:gvar_dist}
\end{figure}

\begin{figure}
\centering
    \includegraphics[width=0.48\textwidth]{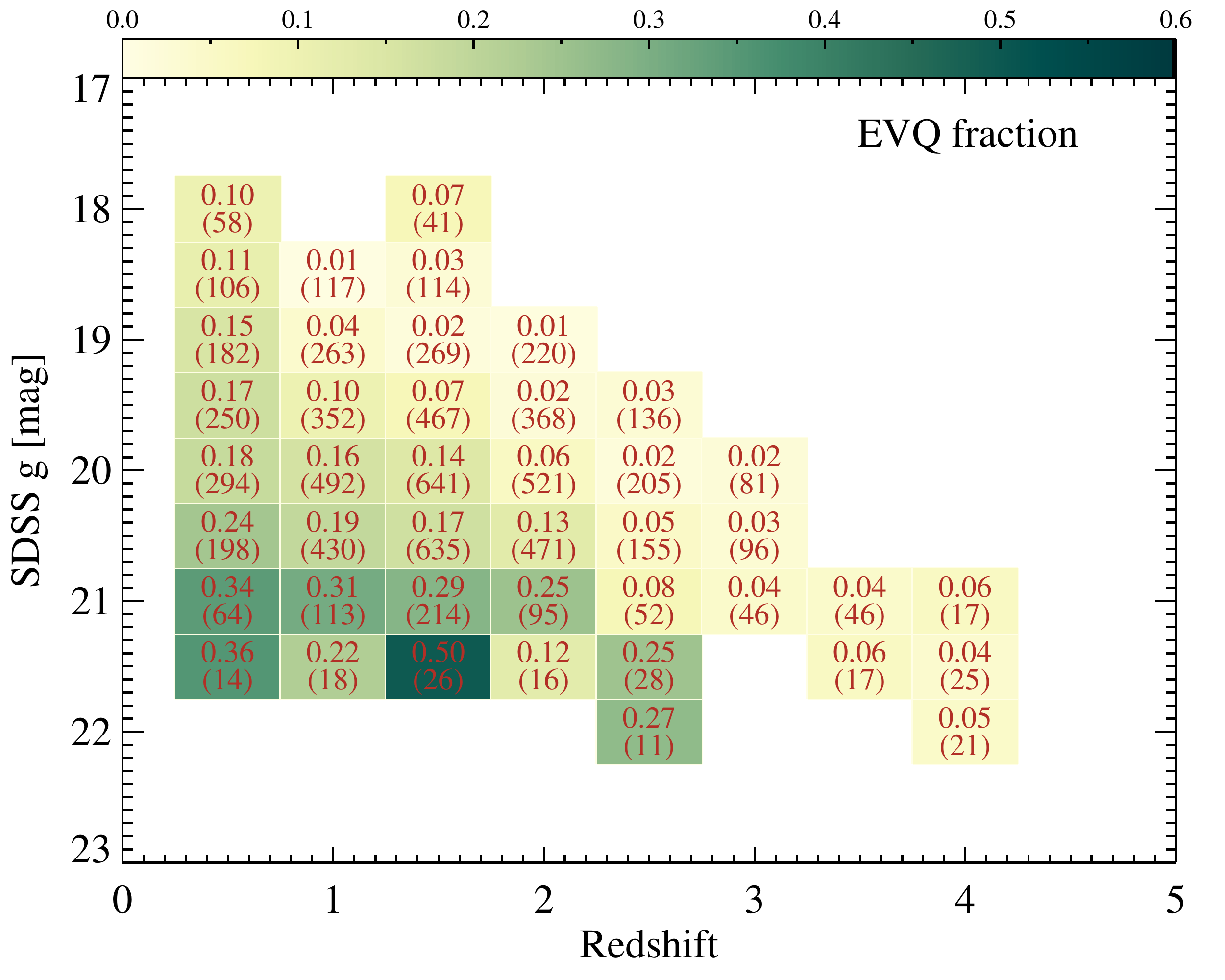}
    \caption{Observed fraction of EVQs as functions of redshift and SDSS $g$ band magnitude. The numbers in the parentheses are the total number of quasars in each bin. Only regions with EVQ detections are shown. The observed EVQ fraction strongly depends on magnitude and redshift. }
    \label{fig:evq_frac}
\end{figure}

\begin{figure}
\centering
    \includegraphics[width=0.48\textwidth]{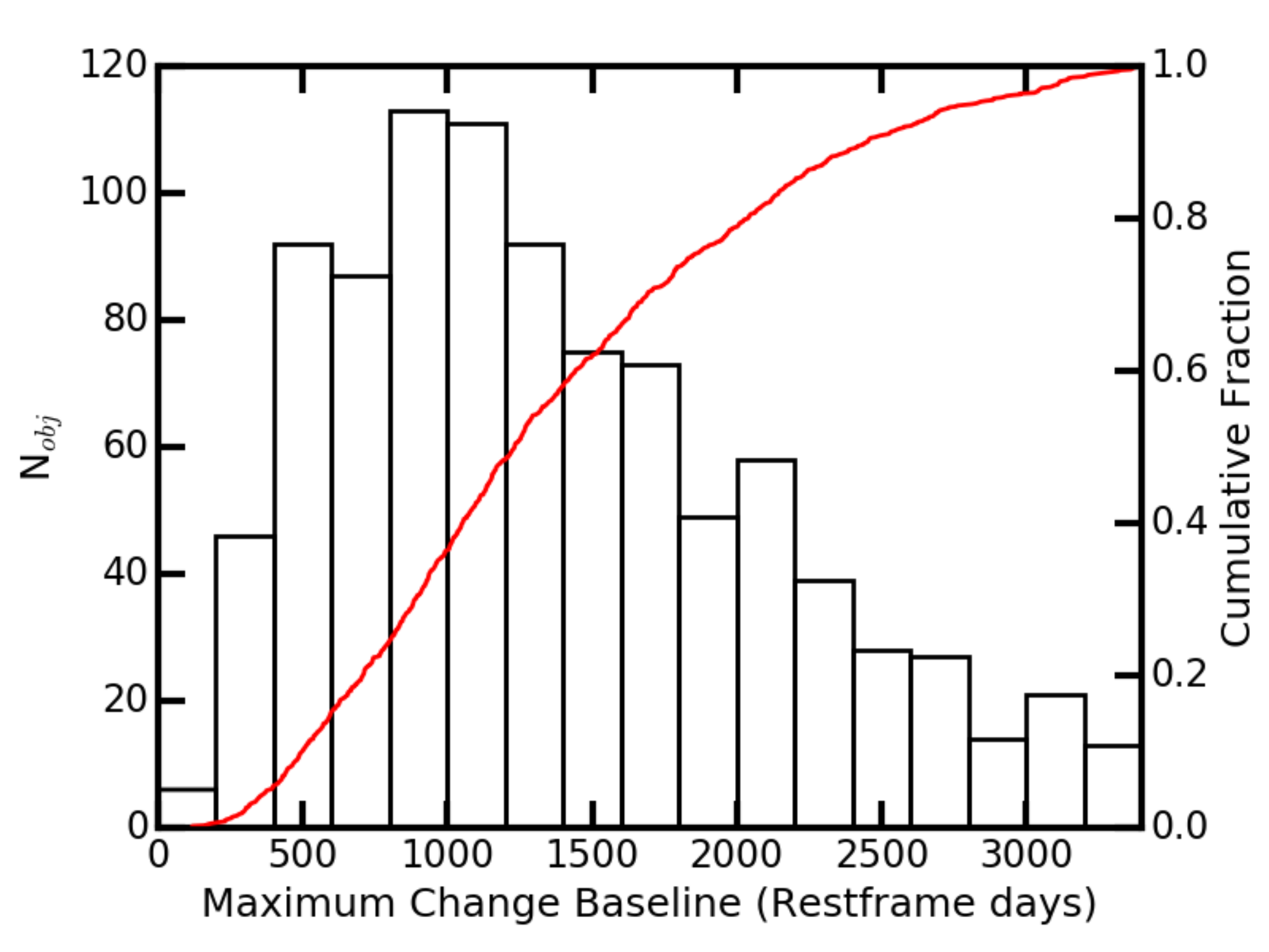}
    \caption{Distribution of rest-frame time separation over which the maximum $g$ band variability is observed for the EVQs. The cumulative distribution is shown with the red line. The apparent dearth of objects beyond $\sim 1500$ days is largely a selection effect due to the time baseline of SDSS$+$DES imaging and gaps in the light curves.}
    \label{fig:baseline_dist}
\end{figure}

According to the criteria described in \S\ref{sec:clq}, we identified 977 objects as EVQs (see Table \ref{tab:sample}). All the previously known CLQs \citep[][]{MacLeod_etal_2016,Ruan_etal_2016} that are within our footprint are recovered. The inclusion of the DES data provided a substantial proportion of these, with 494 EVQs having one extremum in the DES data. Among these 977 EVQs, 373 brightened and 604 dimmed between the two extreme states. This asymmetry between dimmed and brightened EVQs is likely a selection effect: EVQs with much fainter magnitudes in the earlier epochs are more likely to be missed from SDSS, while the DES imaging is much deeper, recovering essentially all SDSS EVQs in the dim state in the DES footprint. Our following statistical analyses also did not find any significant difference in the properties of dimmed and brightened EVQs. 



Fig.\ \ref{fig:gvar_dist} shows the distribution of the maximum $g$-band variation within the time baseline of SDSS$+$DES for all 8640 matched quasars. For most quasars the maximum $g$-band variation is well below 1 magnitude. However, $\sim 10\%$ of the objects show large-amplitude ($>1$ mag) variation during this period, which are selected as EVQs. This overall EVQ fraction is consistent with the results in \citet{MacLeod_etal_2016} on a similar sample of SDSS quasars and with multi-year light curves from SDSS and PanSTARRS 1 \citep{Kaiser_etal_2010}.

The observed EVQ fraction is a strong function of magnitude and redshift, which we demonstrate in Fig.\ \ref{fig:evq_frac}. Fainter quasars have a larger EVQ fraction than brighter quasars, as further shown below. There is also considerable selection bias in the identification of EVQs given the coverage of our light curves, which will be further discussed in \S\ref{sec:selec}.

Fig.\ \ref{fig:baseline_dist} shows the distribution of the (rest-frame) time separations between the epochs of the maximum and minimum $g$-band magnitudes for these EVQs. The drop of objects beyond $\sim 1500$ days is largely caused by the time baseline of SDSS$+$DES imaging and the gaps in the light curves, rather than a real dearth of EVQs at these timescales (see \S\ref{sec:selec}). In future work we will incorporate additional photometric epochs from other surveys, such as PanSTARRS \citep{Kaiser_etal_2010}, to fill the large gap between SDSS and DES, and to acquire additional epochs to extend the observing baseline.  

\begin{figure}
\centering
    \includegraphics[height=0.45\textwidth,angle=-90]{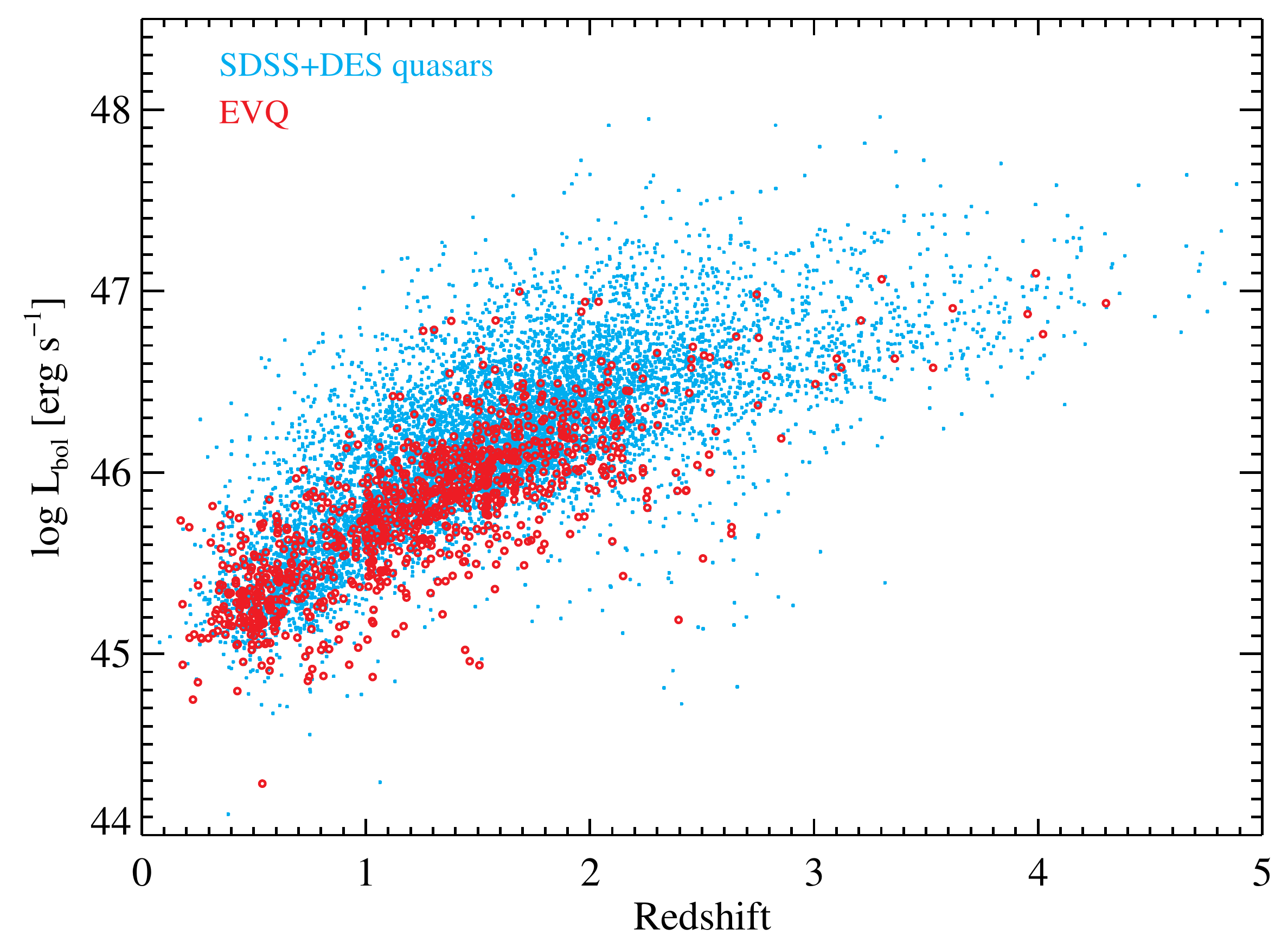}
    \includegraphics[height=0.45\textwidth,angle=-90]{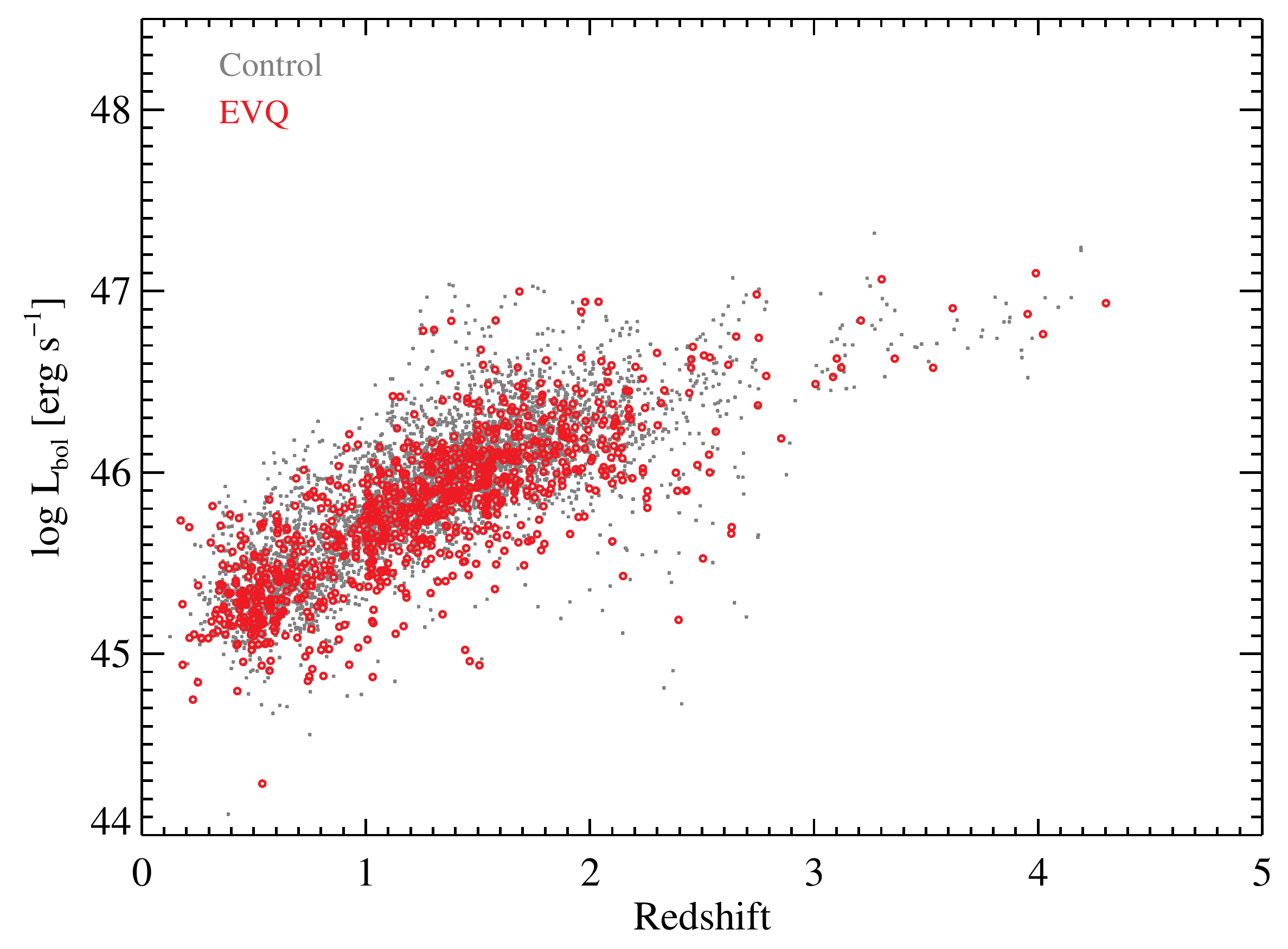}
    \includegraphics[height=0.45\textwidth,angle=-90]{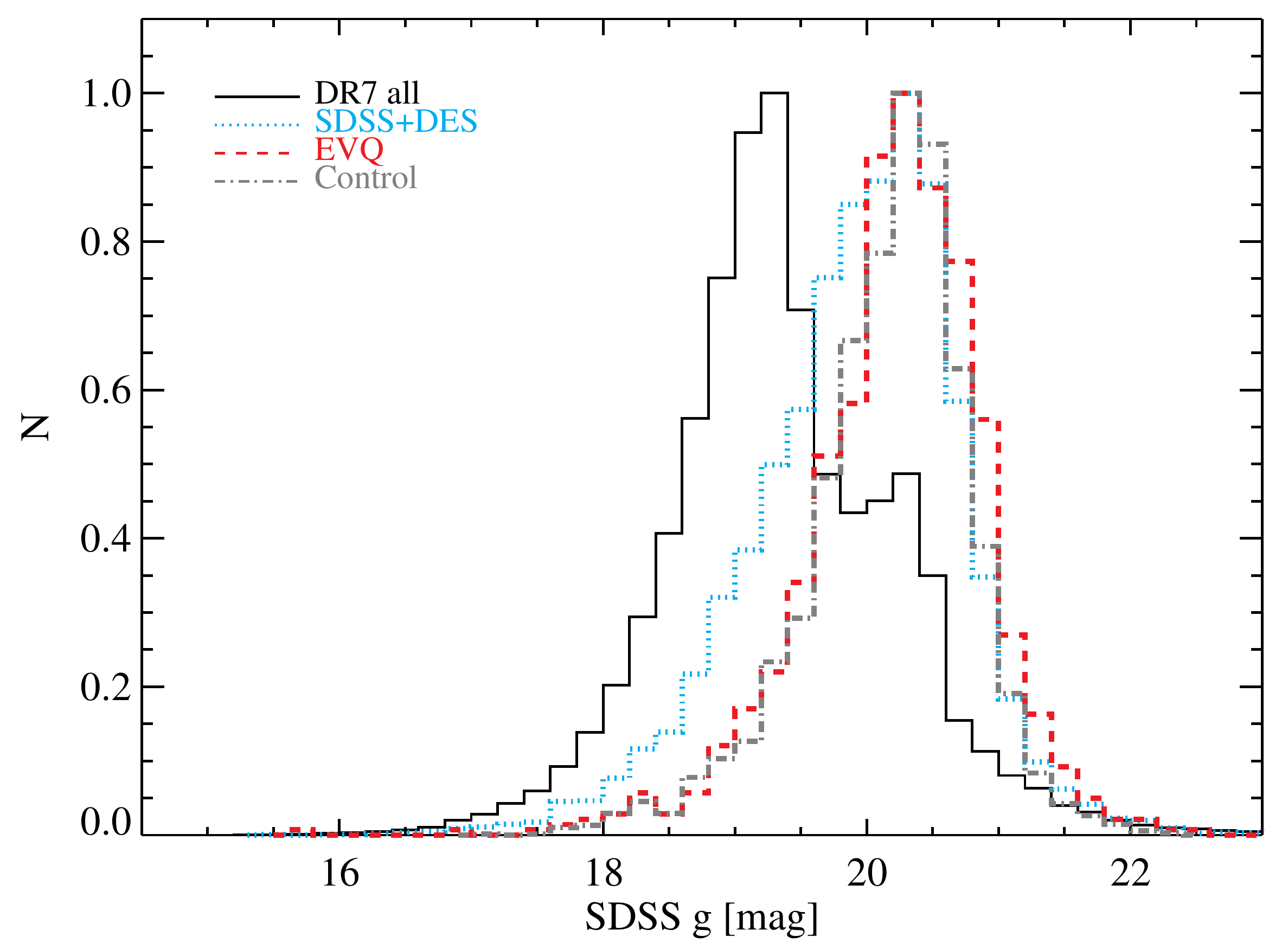}
    \caption{{\it Top}: $L-z$ distributions for the SDSS-DES matched sample (cyan) and the EVQ sample (red), where bolometric luminosities were estimated based on SDSS spectra \citep{Shen_etal_2011}. The EVQs are on average fainter than normal quasars in the SDSS. {\it Middle}: $L-z$ distributions for the EVQ sample (red) and a control sample (gray) matched in redshift and luminosity. {\it Bottom}: Histograms of the $g$ band SDSS magnitudes for various samples. Roughly half of all DR7 quasars were targeted to brighter magnitude limit than the rest of the sample, which led to the shifted peak in the histogram. The EVQs are on average fainter than the SDSS-DES matched sample (also see the top panel). }
    \label{fig:Lz_dist}
\end{figure}

Fig.\ \ref{fig:Lz_dist} (top panel) shows the distribution in the redshift-luminosity plane for the SDSS+DES sample and the EVQ sample, with bolometric luminosities estimated based on SDSS spectra \citep{Shen_etal_2011}. It is apparent from this plot that objects in the EVQ sample have systematically lower luminosities than the DES-matched sample. Since quasar variability decreases with luminosity \cite[e.g.,][]{Vandenberk_etal_2004,Schmidt_etal_2010,Butler_Bloom_2011,MacLeod_etal_2012}, it is reasonable to expect that low-luminosity quasars have a larger probability to show large-amplitude variation over multi-year timescales. 

To reduce confounding factors, and further understand the origin of EVQs, it is important to have a control sample matched in optical luminosities and redshifts. To this end, we created such a sample, which is matched to the EVQ sample in redshift and SDSS $g$ magnitude. The control quasars are drawn from the 9258 quasars in Stripe 82 (most are also in DES footprint, see Table \ref{tab:sample}) and exclude the EVQs. The distribution of the control sample in the $L-z$ plane is shown in the middle panel of Fig. \ref{fig:Lz_dist}. The bottom panel of Fig. \ref{fig:Lz_dist} shows the histogram of $g$ magnitude for the SDSS-DES-matched sample, the EVQ sample and the control sample. Additionally, the black line in Fig. \ref{fig:Lz_dist} shows the distribution of $g$ magnitude in the full DR7Q sample, where most of the objects there were targeted to brighter limiting magnitude \citep[e.g.,][]{Richards_etal_2002a}. We use this control sample for the following analyses.

\begin{table*}
\caption{The EVQ Sample}\label{tab:evq_sample}
\centering
\scalebox{1.0}{
\begin{tabular}{ccccccccccc}
\hline\hline
DR7Q index & RA & DEC & redshift & MJD$_{\rm lo}$ & $g_{\rm lo}$ & $\sigma_{g,lo}$ & MJD$_{\rm hi}$ & $g_{\rm hi}$ & $\sigma_{g,hi}$ & FIRST\_flag \\
 (1) & (2)  & (3) & (4) & (5) & (6) & (7) & (8) & (9) & (10) & (11) \\
\hline
    28         &  0.175101  &  $-0.750386$ & 1.3115 & 52931.22 & 20.900  & 0.040 & 51081.00 & 19.857  & 0.023 &  0 \\
    33         &  0.192320  &  $-0.501993$ & 1.4453 & 54373.38 & 20.636  & 0.100 & 51075.30 & 19.459  & 0.022 &  0 \\
    50         &  0.268872  &  $0.464944$  & 0.5512 & 56546.27 & 21.238  & 0.017 & 51819.36 &  20.189 & 0.026 & 0 \\
    90         &  0.537709  &   $0.098022$ & 2.1447 & 54387.33 & 21.367 &  0.054 & 51819.36 & 20.204 &  0.027 &  0 \\
    97         &  0.579628  &   $0.375815$ & 0.5467 & 53314.21 & 20.722 &  0.036 & 52253.19 & 19.699 &  0.035 &  0 \\
\hline
\hline\\
\end{tabular}
}
\begin{tablenotes}
      \small
      \item NOTE. --- The sample of selected extreme variability quasars and their basic properties. Column (1) is the index of the object in the DR7 quasar catalog of \citet{Shen_etal_2011}. Columns (5)--(10) are the MJD, $g$ magnitude and error for the faintest and brightest epochs in the combined SDSS and DES light curves. Column (11) indicates if the quasar is detected in the FIRST radio survey (1 or 2), undetected (0) or outside the FIRST footprint ($-1$), and is equivalent to the `` FIRST\_FR\_TYPE '' column in the \citet{Shen_etal_2011} catalog. The full table is provided in machine-readable format available online. 
\end{tablenotes}
\end{table*}

\begin{figure}
\centering
    \includegraphics[width=0.45\textwidth]{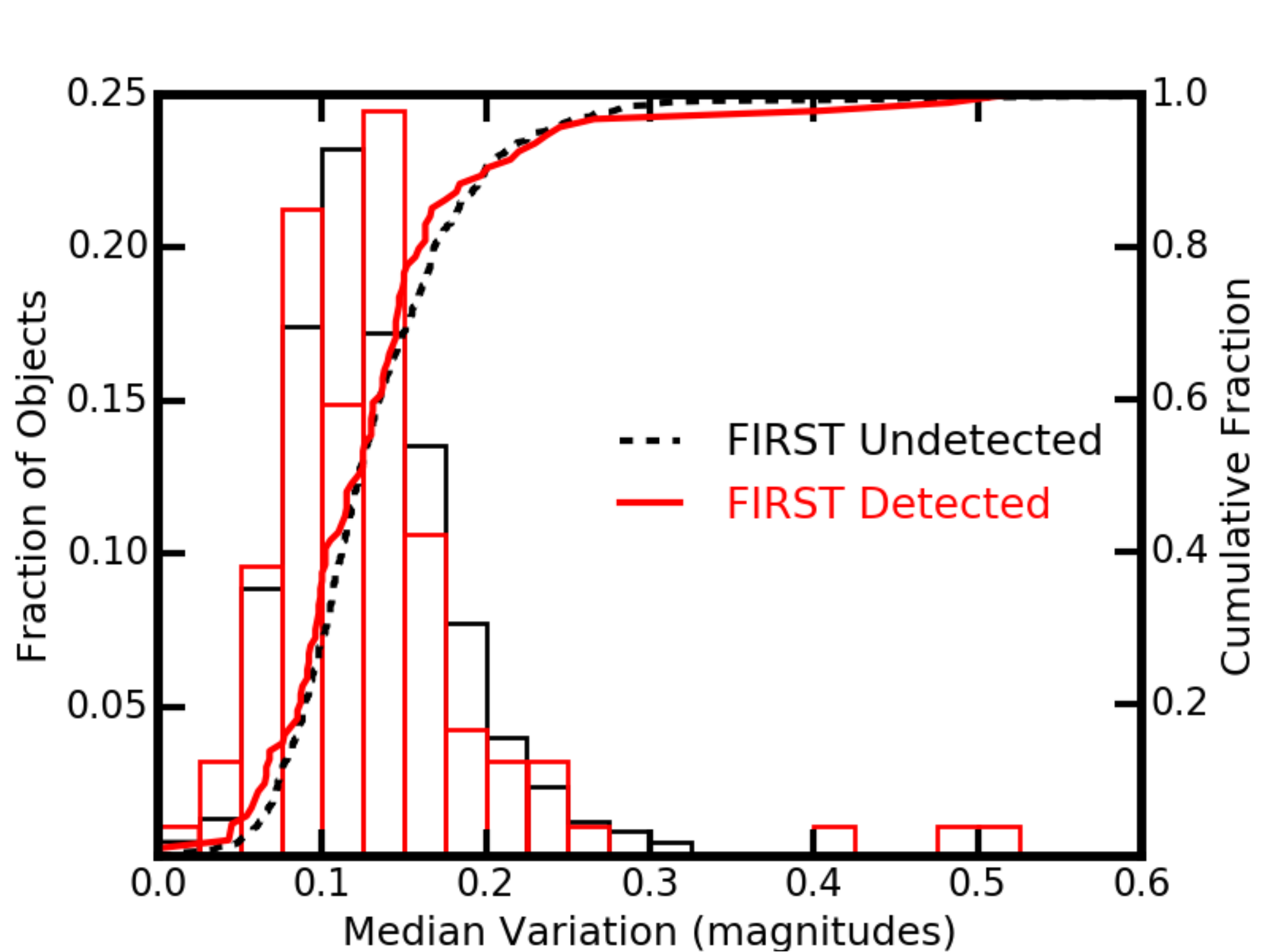}    
    \caption{Distributions of short-term variability (see definition in \S\ref{sec:multi}) for the FIRST-detected (red) and undetected (black) EVQs. The maximum $g$-band magnitude change within a 90-day window for each light curve was used as the metric for this short-term variability. Histograms showing the distribution and the cumulative fractions are plotted.}
    \label{fig:OVV_var}
\end{figure}

\begin{figure*}
\centering
    \includegraphics[width=0.9\textwidth]{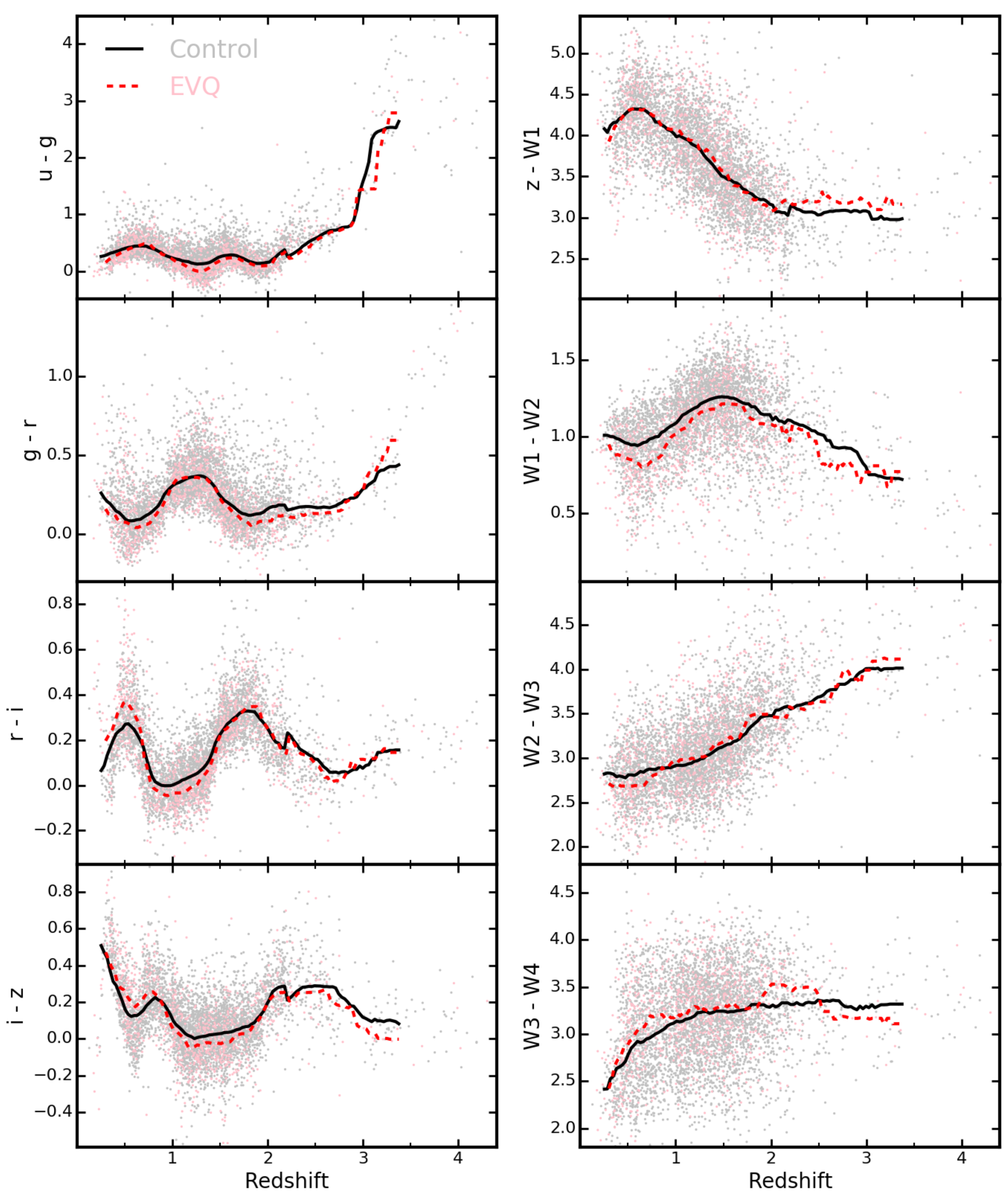}    
    \caption{Redshift tracks of colors for EVQs (pink) and the control sample (cyan). The running median (with a full window size of $\Delta z=0.25$ at $z<2.2$ and $\Delta z=0.8$ at $z>2.2$), relative to redshift, is plotted for the EVQs with a dashed line and for the control sample with a solid line. The optical magnitudes were taken from SDSS and the mid-infrared magnitudes (W1-W4) were taken from the Wide-field Infrared Survey Explorer \citep[WISE;][]{Wright_etal_2010}.} 
    \label{fig:colorplots}
\end{figure*}

\begin{figure*}
\centering
    \includegraphics[height=0.9\textwidth,angle=-90]{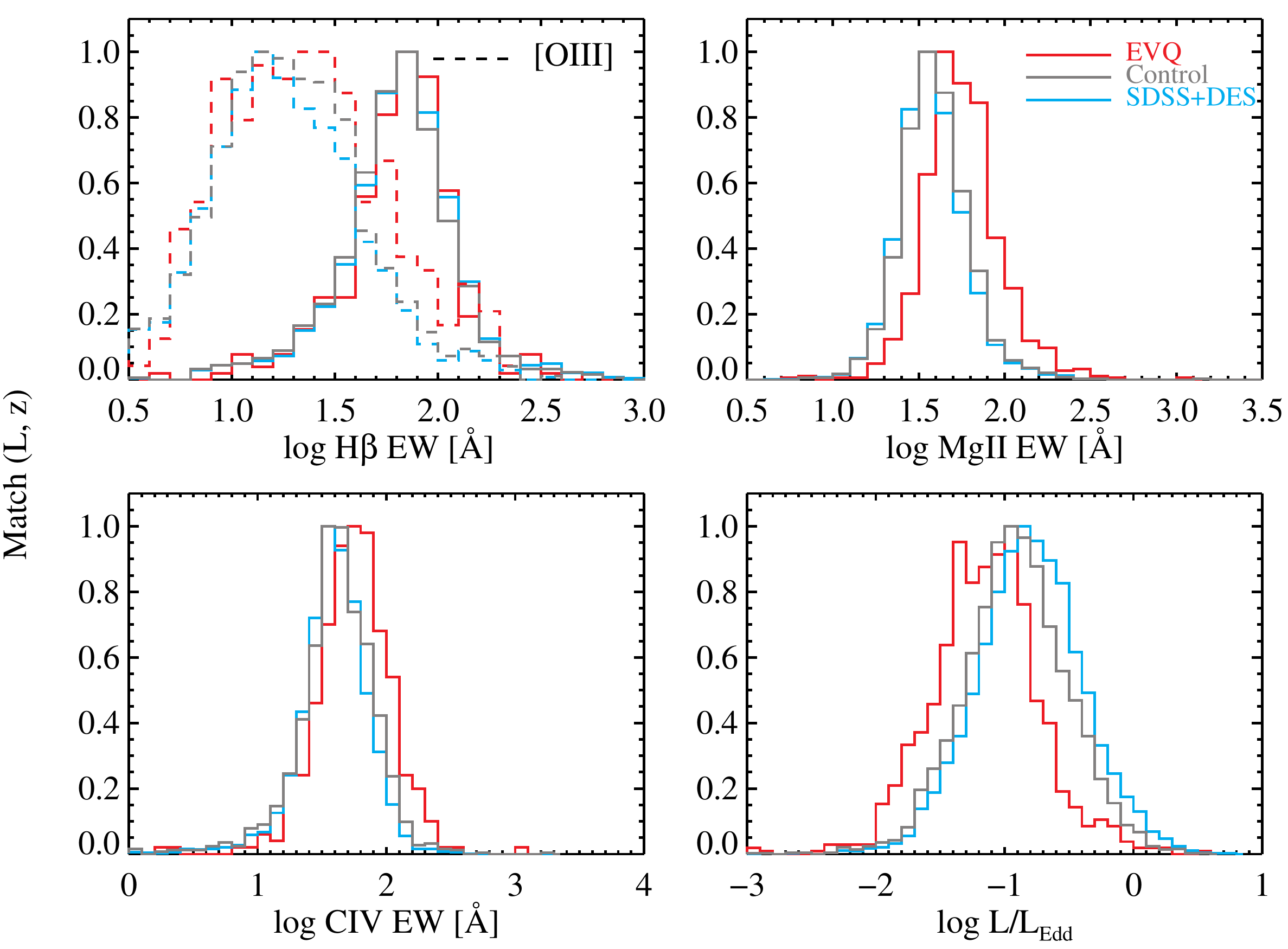}    
    \caption{Emission line strength and Eddington ratio distributions for various samples. EVQs have stronger \MgII\, \CIV\ and \OIII\ lines (i.e., larger EWs) than normal control quasars matched in redshift and luminosity. In addition, EVQs have on average lower Eddington ratios than the control sample. KS tests show that the difference between the EVQ and control sample distributions for EW(\MgII), EW(\CIV), and the Eddington ratio differ by $>6\sigma$, while the distributions for EW(\OIII) differ by $>95$\%.}
    \label{fig:EW_comp}
\end{figure*}

\begin{figure}
\centering
    \includegraphics[width=0.45\textwidth]{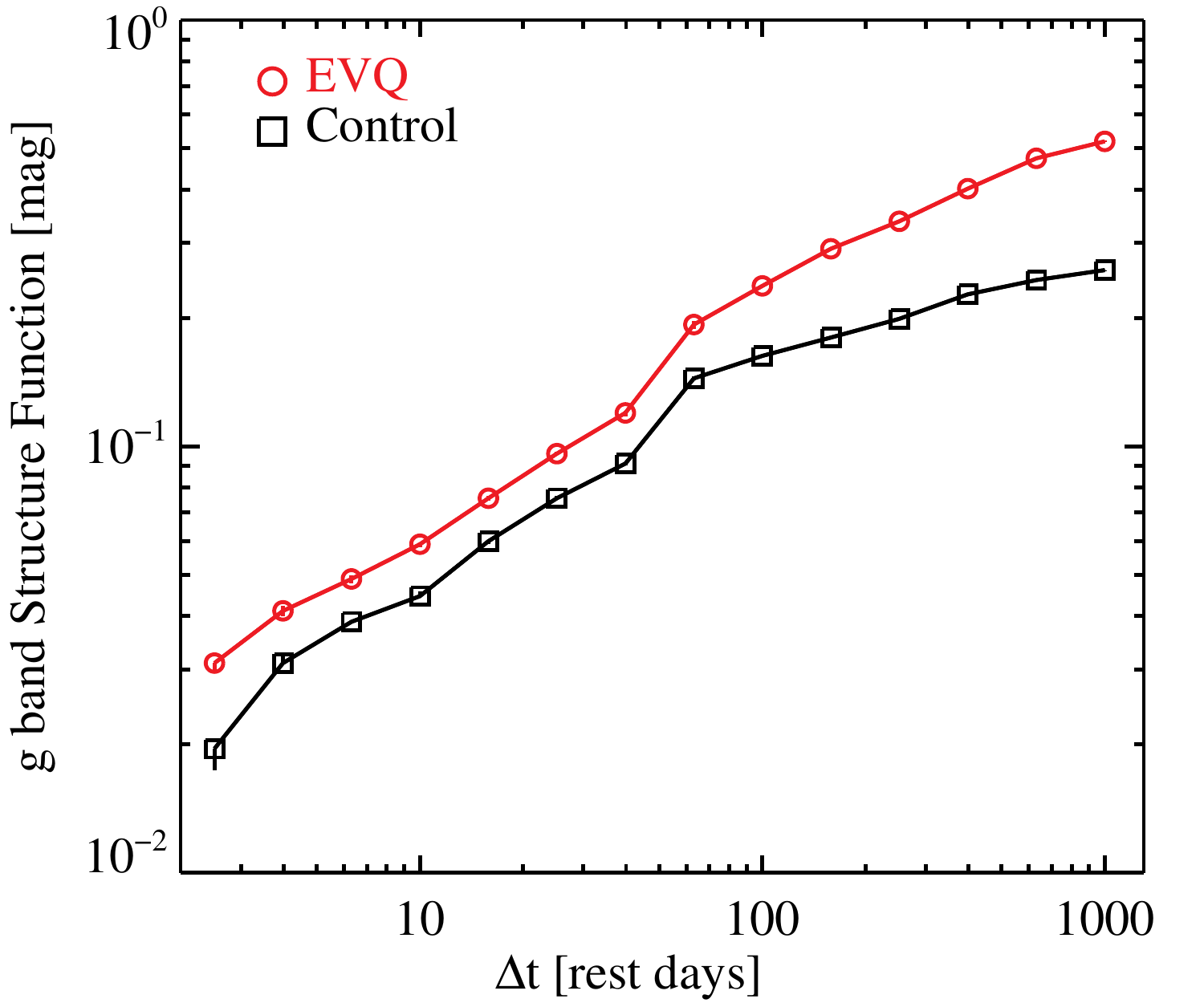} 
    \includegraphics[width=0.45\textwidth]{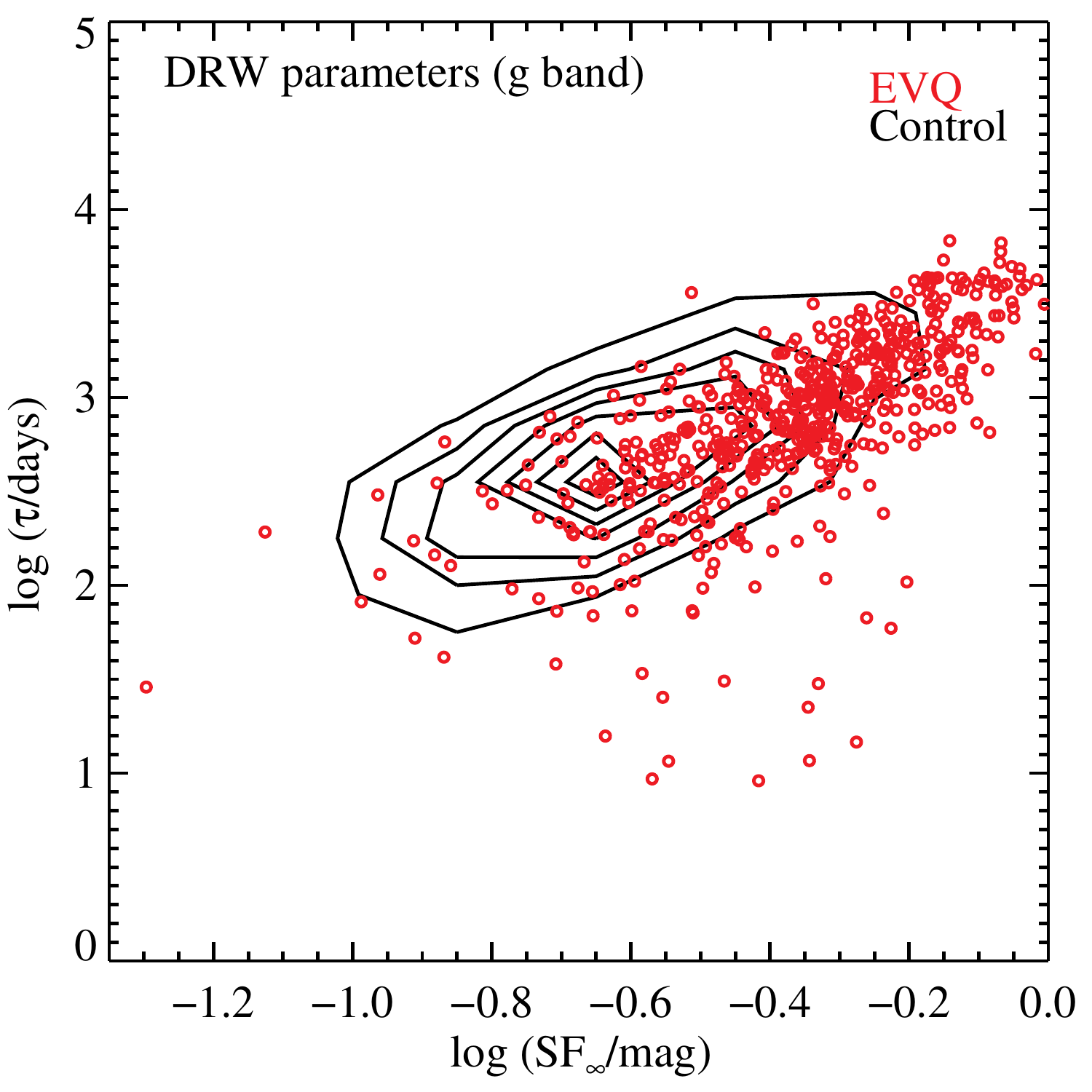}
    \caption{{\it Top}: $g$ band ensemble structure functions computed from the SDSS Stripe 82 light curves, for the control sample and the EVQ sample. EVQs have a larger variability amplitude than control quasars (matched in redshift and luminosity) at all timescales from days to years. {\it Bottom}: DRW model parameters from \citet{MacLeod_etal_2010} for EVQs and control quasars. As expected, EVQs have larger long-term variability amplitude ${\rm SF}_{\infty}$. }
    \label{fig:sf}
\end{figure}

\subsection{Multi-wavelength properties}\label{sec:multi}


We use the radio properties from the FIRST survey \citep{White_etal_1997} as compiled in \citet{Shen_etal_2011} to examine the difference in EVQs and the control sample. Restricting to quasars within the FIRST footprint, there are {93/964 (9.6\%) EVQs and 169/3326 (5.1\%) control quasars detected in FIRST. Thus EVQs are twice as likely to be a radio-loud quasar as normal quasars. 

To investigate the possibility that the large optical variability observed in some FIRST-detected EVQs is due to blazar activity, which occurs on shorter time scales than for typical quasars, we quantify the short-term variability of each EVQ by measuring the maximum magnitude change within a leading 90-day window of each epoch; the median of these maximum magnitude changes for all epochs in the light curve is taken as the metric for the short-term variability of that particular object. Fig.\ \ref{fig:OVV_var} shows the distribution of this short-term variability for the FIRST-detected and FIRST-undetected EVQs. There is no significant difference in the distributions of both populations, and a Kolmogorov-Smirnov (KS) test shows there is a $\sim 30\%$ chance that they are drawn from the same distribution. There are only a few FIRST-detected EVQs showing exceptionally large short-term variability, where the extreme optical variability is likely associated with blazar activity. Thus blazar activity is not a significant contaminant to the observed extreme optical variability in our EVQ sample. As discussed in \S\ref{sec:disc1}, EVQs are generally lower-Eddington ratio quasars, which is consistent with them having a larger radio-loud fraction \citep[e.g.,][]{Ho_2002}.

We further compare the optical and infrared colors of EVQs and the control quasars in Fig.\ \ref{fig:colorplots}. There is no significant difference in the colors of EVQs compared to the control quasars. 

In addition, we examine the changes in color for the quasars in Fig.\ \ref{fig:mag_col_plot}. The changes in $g$ band magnitude between their faintest and brightest epochs for all quasars are plotted versus their respective changes in $g-i$, where $g$ and $i$ are measured at the same, or nearly coincident, epoch. In general, quasars become bluer as they become brighter \citep[e.g.,][]{Vandenberk_etal_2004,Bian_etal_2012,Guo_Gu_2016}, which is what we observe. The EVQ sample appears to be a continuation of the less variable quasars, with no distinction between the populations when considering these color changes. This could imply that the same mechanisms that produce lower variability produce extreme variability as well. If obscuration or TDEs were the cause of extreme variability, we may have expected to see a distinct population of color changes at high $|\Delta g|$.

\subsection{Spectral properties}\label{sec:spec}


Using the spectral measurements from the catalog in \citet{Shen_etal_2011}, we examine the emission line properties in the EVQ sample and compare to those in the control sample. As mentioned earlier, these single-epoch spectral measurements are random representations during the SDSS$+$DES baseline, and probe the average properties of EVQs and control quasars. 

Fig.\ \ref{fig:EW_comp} compares the rest-frame equivalent width (EW) of several major broad (and narrow) emission lines. We found that while the control sample is matched to the EVQ sample in redshift and luminosity, there are significant differences in the emission line strengths between the control sample and the EVQ sample. On average, the UV broad lines (\MgII\ and \CIV) and high-ionization narrow lines (\OIII) of the EVQs have larger EWs than those of the quasars matched in luminosity and redshift. These differences suggest that additional parameters, other than luminosity, are causing the difference in their emission line strength.  

We argue that the different emission line properties in EVQs can be explained by the Eddington ratio $L/L_{\rm Edd}$, where $L_{\rm Edd}$ is the Eddington luminosity of the black hole. To test this hypothesis, we plot the distributions of Eddington ratios from \citet{Shen_etal_2011} for different samples in Fig.\ \ref{fig:EW_comp} (lower-right panel), where the BH mass is estimated using the so-called single-epoch virial BH mass estimators \citep[for a recent review, see][]{Shen_2013}. Most of the BH masses were estimated based on the broad \hbeta\ and \MgII\ lines at $z<1.9$, with the remaining objects at $z>1.9$ estimated with the less reliable broad \CIV\ line \citep[we refer the reader to][for a detailed discussion on the caveats of BH masses estimated with different lines]{Shen_2013}. Bearing in mind the large ($\sim 0.5$ dex) systematic uncertainties in these Eddington ratio estimates, there is evidence that EVQs have lower Eddington ratios than quasars in the control sample. 

We will further discuss the connection between line strength and Eddington ratio in \S\ref{sec:disc2}.

\subsection{Variability properties}


There are various ways to characterize the variability properties of quasars. The structure function \citep[SF, e.g.,][and references therein]{Kozlowski_2016} describes the typical variability amplitude between epochs separated by a certain timescale. This is a purely empirical approach and does not have the ambiguities of model fitting and interpretation \citep[e.g.,][]{Kozlowski_2016}. 

Fig.\ \ref{fig:sf} (top) shows the $g$ band ensemble SFs of the EVQ sample and the control sample as a function of rest-frame time separation. EVQs are more variable than normal quasars matched in luminosity and redshift at all timescales, and the excess variability of EVQs is more significant on multi-year timescales, a reflection of them showing extreme ($>1$ mag) variability over such long timescales.

Quasar variability can also be modeled as a stochastic process. In recent years, the damped random walk \citep[DRW, e.g.,][]{Kelly_etal_2009,Kozlowski_etal_2010} model has been successfully applied to model optical light curves of quasars \citep[e.g.,][]{MacLeod_etal_2010}. Following the convention in \citet{MacLeod_etal_2010}, the DRW model has two independent parameters, the damping timescale $\tau$, and the long-term variability amplitude ${\rm SF}_{\infty}$ (i.e., the asymptotic structure function at very long time separations). Fig.\ \ref{fig:sf} (bottom) shows the distributions of the DRW parameters for the EVQ and the control samples, where the values are taken from \citet{MacLeod_etal_2010} measured for all SDSS Stripe 82 quasars. Consistent with the SF results, the DRW modeling also shows larger long-term variability for EVQs compared to normal quasars matched in luminosity and redshift. 

\section{Discussion}\label{sec:disc}

\subsection{EVQs are low Eddington ratio systems}\label{sec:disc1}

The findings in \S\ref{sec:clq} and in particular the spectral properties presented in \S\ref{sec:spec} led to the simple interpretation that EVQs are the low-Eddington-ratio subset of the general quasar population. Past quasar variability studies based on structure function or stochastic models such as the damped random walk (DRW) model have shown that quasars with higher Eddington ratios vary less than those with lower Eddington ratios \citep[e.g.,][]{Ai_etal_2010}. By extension, then, low Eddington ratio objects are also more likely to display large-amplitude variation over multiple years, as observed here.

Fig.\ \ref{fig:Edd_gvar} displays the relation between estimated Eddington ratio and the maximum variability over the course of SDSS$+$DES. Note that individual Eddington ratio estimates are quite uncertain \citep[e.g.,][]{Shen_2013}, and so one should look at the average Eddington ratio (e.g., the cyan points in Fig.\ \ref{fig:Edd_gvar}) as a function of the maximum $g$-band variability. As expected, the average Eddington ratio decreases as the maximum $g$-band variability increases. 

We further illustrate the role of Eddington ratio in driving the extreme variability in EVQs in Fig.\ \ref{fig:EV1}, where we plot the distribution in the broad \hbeta\ width versus the normalized optical \FeII\ strength $R_{\rm FeII}\equiv {\rm EW_{FeII,4434-4684{\textrm \AA}}/EW_{H\beta}}$, for the low-$z$ subset of our sample with \hbeta\ coverage. The sequence from left to right in this plot (i.e., increasing \FeII\ strength), is known as the Eigenvector 1 \citep[e.g.,][]{Boroson_Green_1992}, and is believed to be driven by Eddington ratio \citep[e.g.,][]{Boroson_Green_1992,Sulentic_etal_2000,Boroson_2002,Shen_Ho_2014}. The EVQ sample lies systematically towards the left of the distribution, compared to the control sample and the full DR7Q sample, which is consistent with the above interpretation that EVQs are low-Eddington ratio systems. 

The different properties of EVQs compared to normal quasars suggest that the eclipsing cloud scenario \citep[e.g.,][]{Risaliti_etal_2009} and the tidal disruption event scenario \citep[e.g.,][]{Merloni_etal_2015} cannot account for the majority of EVQs, and by extension, CLQs, unless there is certain correlation between the rates of the eclipsing or TDE events and the Eddington ratio of the quasar. 

\begin{figure}
\centering
    \includegraphics[height=0.45\textwidth,angle=-90]{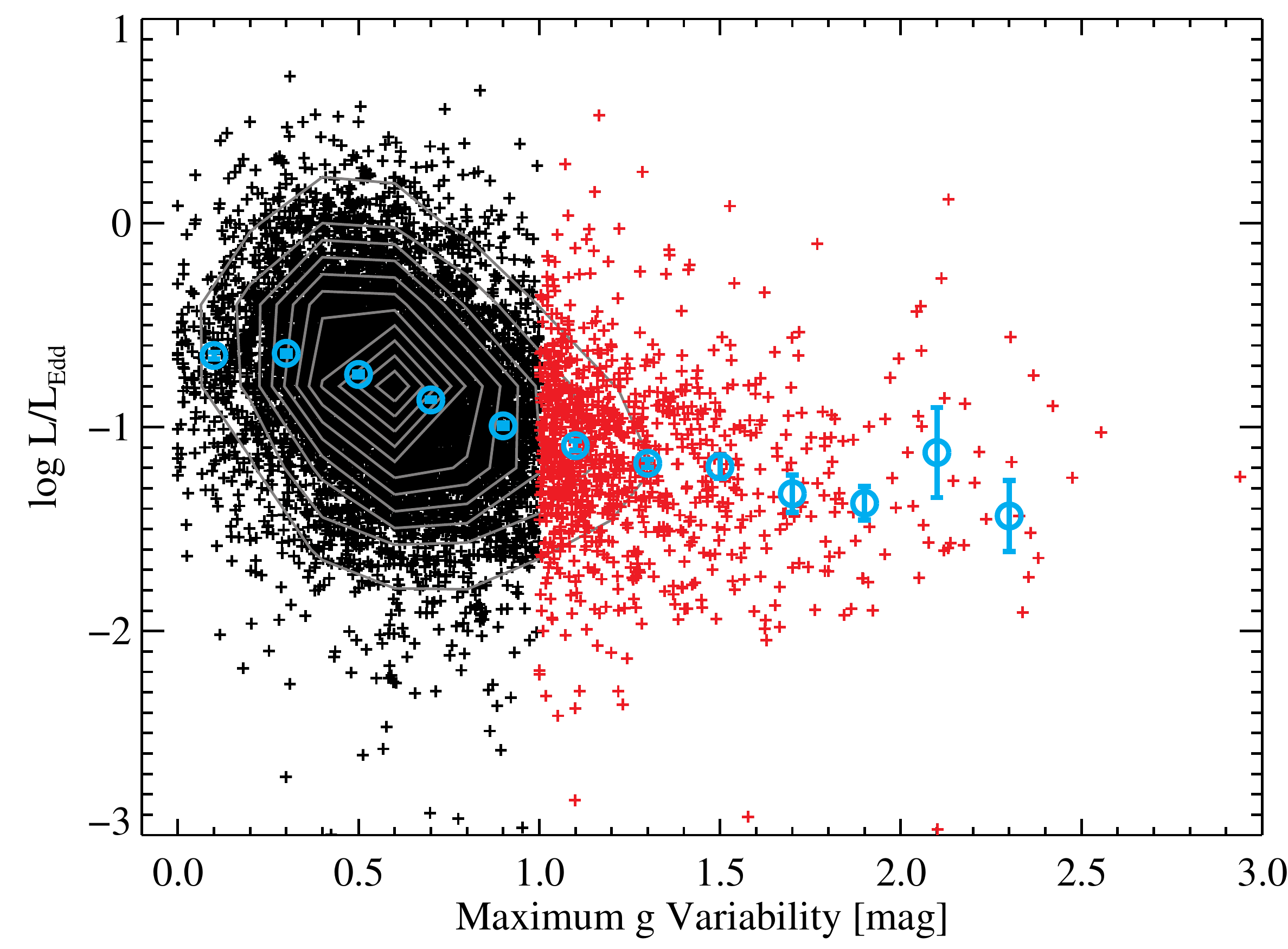}    
    \caption{Correlation between maximum $g$-band variability and Eddington ratio estimated from SDSS spectrum. There is a general trend of decreasing Eddington ratio when the maximum $g$-band variability increases. The red points are the selected EVQs, and the cyan points are the median Eddington ratio in each bin of maximum $g$ variability. }
    \label{fig:Edd_gvar}
\end{figure}

\begin{figure}
\centering
    \includegraphics[height=0.45\textwidth,angle=-90]{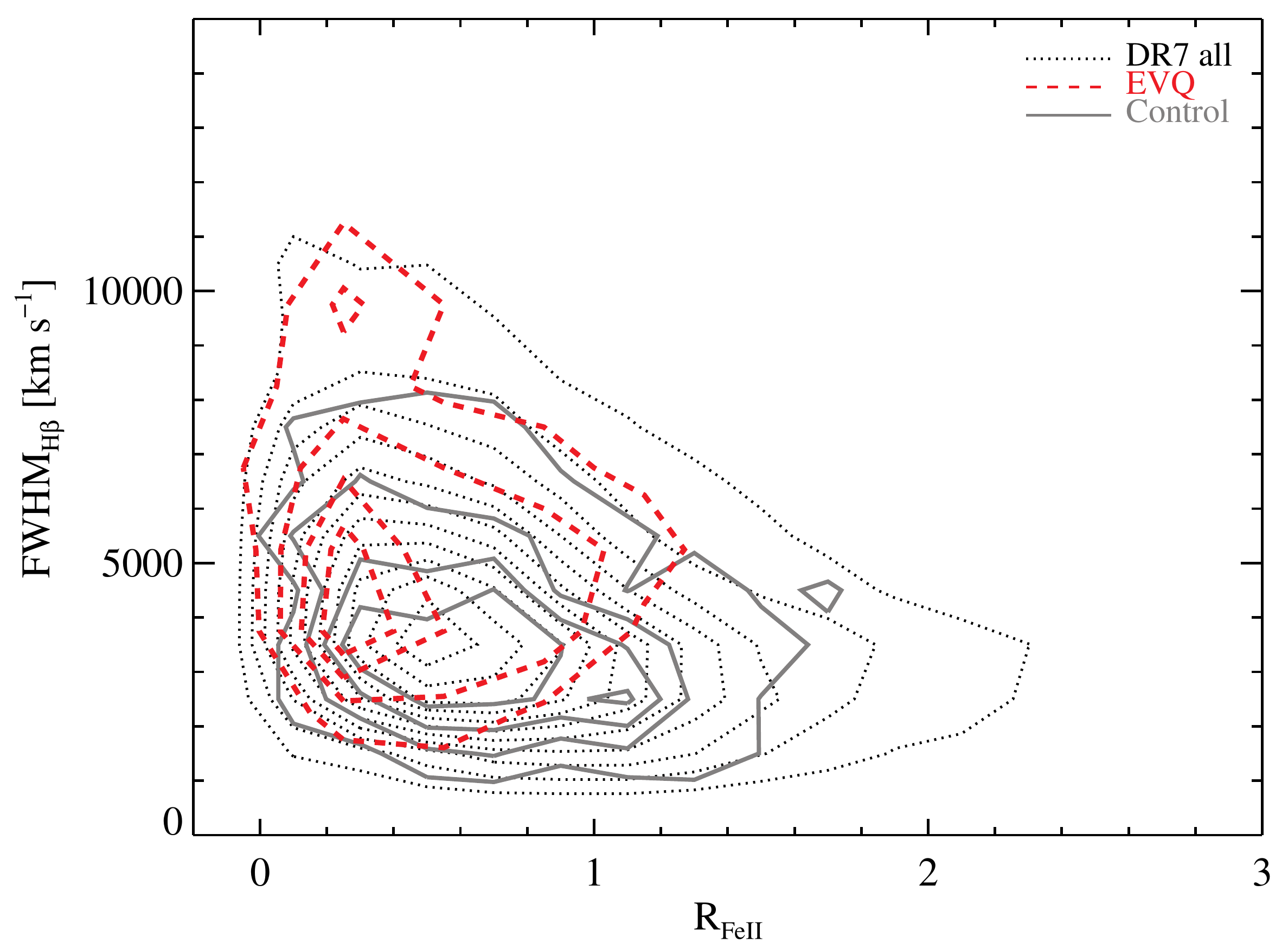}    
    \caption{Distribution of quasars in the broad \hbeta\ FWHM versus optical \FeII\ strength $R_{\rm FeII}\equiv {\rm EW_{FeII,4434-4684{\textrm \AA}}/EW_{H\beta}}$. EVQs have weaker \FeII\ strength compared with the control sample and all DR7 quasars, consistent with them being lower-Eddington ratio systems (see discussion in \S\ref{sec:disc1}).}
    \label{fig:EV1}
\end{figure}

\subsection{Connection to weak-line quasars}\label{sec:disc2}

The correlation between line strength and the Eddington ratio of quasars has been the focus of many studies in recent years. For example, \citet[][]{Dong_etal_2009} showed that there is a strong anti-correlation between the \MgII\ EW and Eddington ratio, mostly resulting from a correlation between \MgII\ EW and broad \MgII\ FWHM \citep[also see Fig. 13 in][]{Shen_etal_2011}, and secondly from an anti-correlation between \MgII\ EW and quasar continuum luminosity. Since the broad \MgII\ FWHM and continuum luminosity are combined to provide an estimate of the virial BH mass \citep[e.g.,][]{Shen_2013}, an anti-correlation between \MgII\ EW and $L/L_{\rm Edd}$ emerges. 

Radio-quiet quasars with significantly weaker broad emission lines, termed ``weak line quasars'' \citep[WLQs, e.g., ][]{Fan_etal_1999, Plotkin_etal_2008, Plotkin_etal_2010, Diamond_Stanic_etal_2009, Shemmer_etal_2010,Shemmer_Lieber_2015}, are often X-ray weak compared to quasars with normal broad-line strength \citep[e.g.,][]{Wu_etal_2011, Luo_etal_2015}. The latter studies proposed a scenario where there is a geometrically-thick ``shielding gas'' in the innermost region of the accretion disk \citep[as motivated by earlier ideas in, e.g.,][]{Madau_1988,Leighly_2004}, which blocks the hard ionizing continuum from the inner disk (and the X-ray flux from the hot corona immediately surrounding the BH) from reaching the BLR (and the \OIII\ narrow-line region), resulting in the observed weak line emission. Depending on the orientation of the system, the X-ray emission is either visible to the external observer when viewed at high inclinations (X-ray normal), or blocked by the shielding gas along the light-of-sight to the observer (X-ray weak). In any case, the BLR receives fewer ionizing photons and hence the broad line strength (in particular the strength of high-ionization lines) is reduced. 

A plausible origin for this shielding gas is a geometrically thick inner accretion disk, such as in the slim disk model \citep[e.g.,][]{Abramowicz_etal_1988,Wang_etal_2014}. The slim disk model is naturally connected to Eddington ratio: when the Eddington ratio is high ($L/L_{\rm EDD}\gtrsim 0.3$), optically thick advection becomes important and the slim accretion disk becomes a more appropriate solution than the standard thin disk model \citep[][]{Shakura_Sunyaev_1973}. Alternatively, in the global simulations of high accretion-rate disks by \citet{Jiang_etal_2014}, the disk is unlikely to maintain a thin configuration throughout and may be puffed up in the inner region, which naturally provides the required shielding gas. Although these theoretical studies generally focused on high Eddington ratio systems, we speculate that there is a continuous trend in the relative importance of this shielding gas as accretion rate increases, which then naturally leads to the observed correlation between broad-line strength and Eddington ratio. The appeal of this scenario is that it can also qualitatively explain other observed trends in quasar properties, such as the dependence of the shape and blueshift of the \CIV\ line on luminosity and Eddington ratio \citep[e.g.,][]{Richards_etal_2011}.

The EVQs studied here have stronger broad emission lines (i.e., larger EWs) when compared to the control sample matched in quasar continuum luminosity. In the context of WLQs discussed above, EVQs should have on average lower Eddington ratios, fully consistent with the implications from their variability and spectral properties.


Finally, we comment on the potential effect of orientation. It is often tempting to attribute weak line strength to orientation effects, where the system is viewed more pole-on and hence the continuum flux from a geometrically thin, optically thick disk is larger than that viewed from a more edge-on position, and the broad line EW is thus reduced. This interpretation cannot explain the observed correlation between line EWs and EVQs (e.g., the variability properties are intrinsically different for these EVQs with stronger broad lines). Furthermore, recent work comparing the line EWs in normal and broad-absorption-line quasars also suggests that there might be some ambiguities in using line EWs as an orientation indicator \citep[][]{Matthews_etal_2017}. Our results, while only based on a specific subset of quasars showing extreme variability, support the idea that the weakness of the broad lines is mostly intrinsic to the properties of the quasar, rather than mostly due to an orientation effect. 

\subsection{{Timescales and frequency of extreme variability}}\label{sec:selec}

\begin{figure}
\centering
    \includegraphics[width=0.48\textwidth]{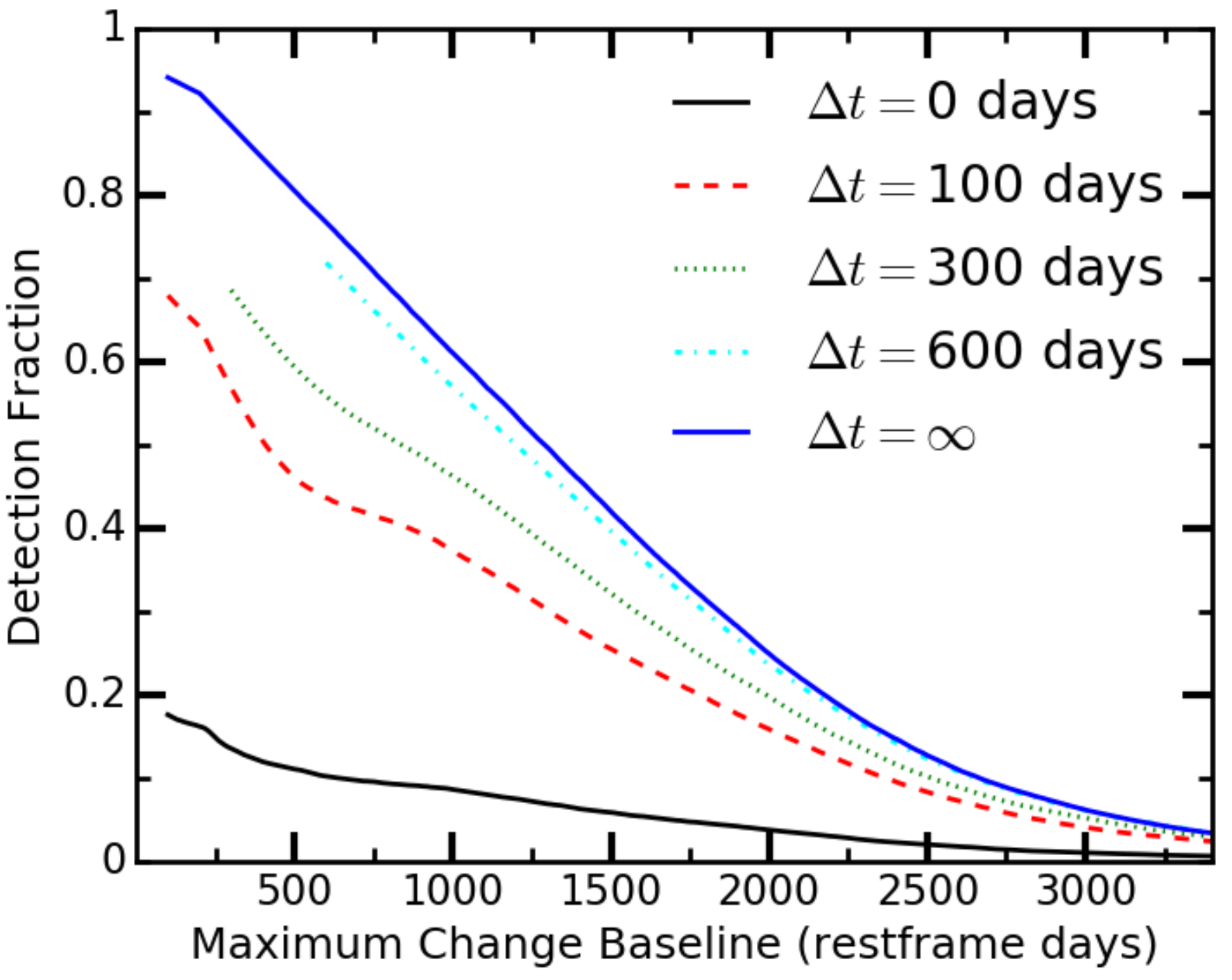}   
    \caption{Estimated detection fraction of EVQs as a function of the rest-frame timescale of the extreme variability, based on the simulations described in \S\ref{sec:selec}. The blue solid line represents the maximum detectability, where only the baseline of the observations matter. The other lines correspond to simulations where detections were determined when a ``buffer'' region overlapped the mock observing seasons, representing a duration of the extreme states ($\Delta t$). These demonstrate the additional impact from gaps in the light curves. Shorter values of $\Delta t$ will lead to a larger impact of light curve gaps on the detectability. }
    \label{fig:det_frac}
\end{figure}

We now examine the frequency of EVQs and the timescales over which extreme variability can occur. We use a simple model to estimate the intrinsic fraction of EVQs, given the observed fraction. This model has several assumptions, and does not address potential correlations among different properties of EVQs and their dependencies on the extreme variability timescale. Nevertheless, we use this exercise as a rough guideline to understand the frequency of EVQs.  

First, we assume that the extreme variability (i.e., defined as $|\Delta g|>1$) can occur with a rest-frame separation of $\Delta T$, and each of the two extreme states has a rest-frame duration of $\Delta t$. We also want to quantify the number of EVQs within our observing baseline only, since extreme episodes at some arbitrary time in the past or future (e.g., 50 years before our first observation) are not meaningful for our analysis. Our observing baseline provides a useful timeframe within which to analyze the population. We therefore require at least one extreme epoch of each mock EVQ to be within our overall observing baseline.

Given these assumptions on the timescale and duration of extreme variability, an EVQ will be observed if: (a) both extreme epochs are covered by the total baseline of the observation, and (b) neither epoch is lost due to large gaps in the light curve coverage. If the duration of the extreme states $\Delta t$ is longer than the gaps in the light curve, then the detection probability is simply one minus the ratio between the timescale of the extreme variability $\Delta T$ and the baseline of the observation. For example, if the baseline of the observation is 10 years and the timescale of the extreme variability is 5 years (both in rest-frame), the detection probability would be 50\%. On the other hand, if $\Delta t$ is shorter than the seasonal gaps, the EVQ might be lost if the extreme epoch is too deep into the gap. In this sense, $\Delta t$ serves as a ``buffer'' to increase the detectability of EVQs in the presence of gaps in the observations, and results with a very large $\Delta t$ will approach the limiting case of maximum detectability\footnote{For simplicity and self-consistency, we do not consider the effect of the buffer if one or two of the extreme epochs fall outside the baseline of the observation. In other words, the buffer $\Delta t$ will only remedy the impact of gaps in the light curve.}. In this case we designate $\Delta t=\infty$, and consider an EVQ detectable as long as both extreme epochs are within the observational baseline.  

With this simple model, we proceed to estimate the detection fraction as a function of the rest-frame timescale of the extreme variability with simulations. The simulated observed-frame baseline of the observation is $\sim 16$ yrs, with seasonal gaps and one large 4-year gap to mimic the combined SDSS$+$DES observations. We generate a mock sample of EVQs with a flat distribution in $\Delta T$; each of these mock EVQs is assigned a random redshift drawn from the SDSS$+$DES quasar sample and both $\Delta T$ and $\Delta t$ are inflated by $(1+z)$. We then randomly assign one epoch of the two extreme states within the observational baseline, randomly determine if the other epoch occurred earlier or later, and record the mock EVQs that are detected in the simulated observation. The detection fraction is then derived as a function of $\Delta T$, and the results with several different values of $\Delta t$ are shown in Fig.\ \ref{fig:det_frac}. To be self-consistent, for a finite value of $\Delta t$, we do not consider objects with a $\Delta T$ less than the assumed $\Delta t$ value. Since we observed very few EVQs in the actual data with $\Delta T>3400$ days (rest-frame), we limit the comparison to $\Delta T<3400$ days.  

The detection fractions in Fig.\ \ref{fig:det_frac} suggest that we are preferentially missing EVQs with large $\Delta T$ values. This is expected given the limited time baseline of our observation. However, Fig.\ \ref{fig:det_frac} also suggests that $\Delta t$ cannot be too small, otherwise the implied intrinsic fraction of EVQs will be too high to distinguish them from the general population in their properties, as discussed in \S\ref{sec:clq}.

Fig.\ \ref{fig:int_frac} shows the intrinsic EVQ fraction after we correct for the selection incompleteness, for two cases with $\Delta t=\infty$ (left) and $\Delta t=100$ days (right). The overall detection fraction over $0-3400$ rest-frame days as probed by our observations is $\sim 0.3$ and $\sim 0.2$ in the two cases. This result indicates that the intrinsic fraction of EVQs is between $\sim 30-50\%$, much higher than the observed fraction of $\sim 10\%$. We do not consider smaller $\Delta t$ values are possible because in this case EVQs would become the majority and would not distinguish from the control sample in quasar properties, as we demonstrated in \S\ref{sec:clq}.  

After correcting for the detection fraction, Fig.\ \ref{fig:int_frac} indicates that extreme variability can occur over a broad range of timescales, with mild evidence that more EVQs occur over longer timescales. We compared all properties of EVQs in two subsamples divided by the rest-frame timescale of the extreme variability and found indistinguishable results. This suggests that the same mechanism that drives this extreme variability can operate on a variety of timescales.

\begin{figure*}
\centering
    \includegraphics[width=0.48\textwidth]{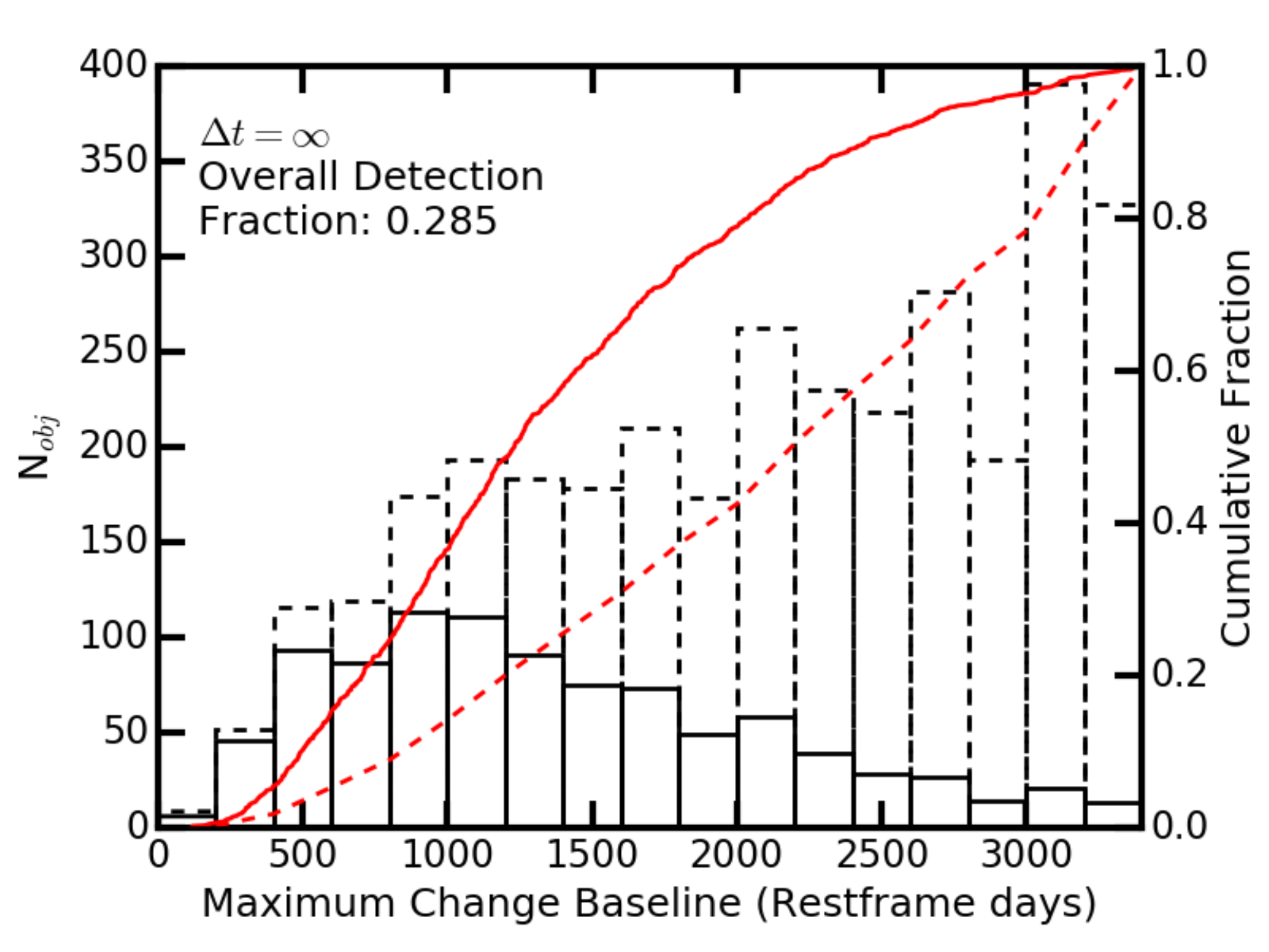} 
    \includegraphics[width=0.48\textwidth]{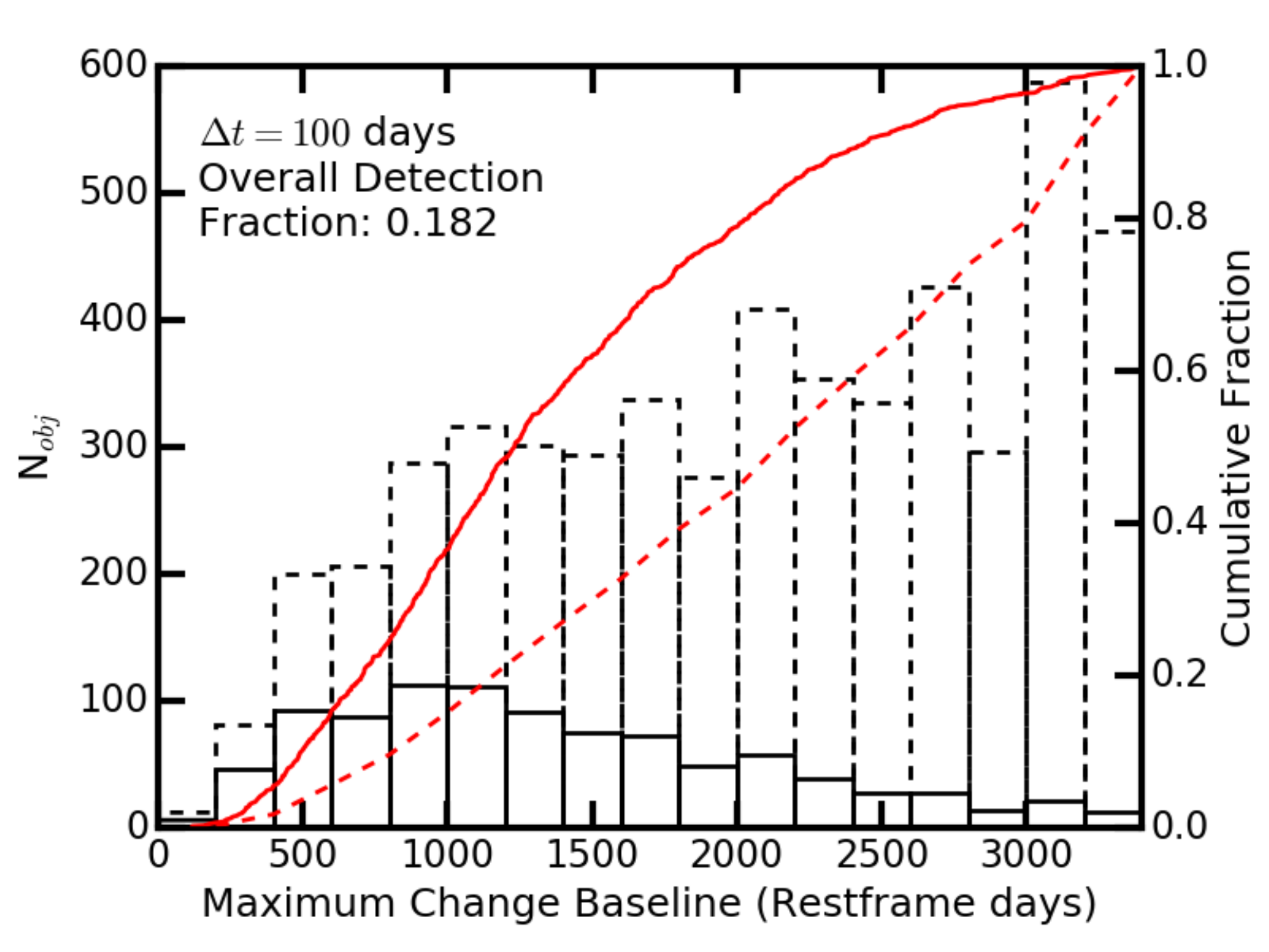}
    \caption{Distribution of timescales of extreme variability. The solid histogram and line shows the observed distribution, and the dashed histogram and line shows the intrinsic distribution after correcting for selection based on the simulations described in \S\ref{sec:selec}. The overall detection fraction over rest-frame $0-3400$ days is marked in the upper-left corner. {\bf Left:} the case of maximum detectability. {\bf Right:} the case with $\Delta t=100$ days. See details in \S\ref{sec:selec}. }
    \label{fig:int_frac}
\end{figure*}

Finally, as a sanity check on the simple model approach above, we perform a different simulation using the damped random walk model for quasar variability \citep[e.g.,][]{Kelly_etal_2009}. We use the DRW parameters for the SDSS Stripe 82 quasars from \citet{MacLeod_etal_2010} to generate stochastic quasar light curves using the DRW prescription over a rest-frame baseline of 6000 days, and identify the faintest and brightest epochs. We then down-sample these light curves using the duration and cadence of our combined SDSS and DES observations, and identify the faintest and brightest epochs in the {\it observed} light curves. We restrict to 7406 Stripe 82 quasars presented in \citet{MacLeod_etal_2010} with reliable DRW fits, which roughly match the parent SDSS$+$DES sample in our EVQ search.  

With the simulated quasar light curves, we identify 362 and 1097 quasars with $|\Delta g|>1$ within a rest-frame baseline of 3400 days as in our real observations, for the ``observed'' and ``intrinsic'' cases, respectively. The distributions of the timescale of the extreme variability in this simulation look similar to those in our simple model approach (e.g., Fig.\ \ref{fig:int_frac}). The derived overall detection fraction (with $\Delta T<3400$ days) is 0.33, again similar to what we found in the above simple model. However, the observed EVQ fraction, $362/7406\approx 5\%$, is significantly smaller than the observed $\sim 10\%$ EVQ fraction. This suggests that the DRW model is not a perfect prescription for describing extreme quasar variability, which may deviate from a random Gaussian process. Relaxing the magnitude cut to $|\Delta g|>0.8$, we identify 895 and 2079 quasars for the ``observed'' and ``intrinsic'' cases, respectively, which are more in line with the actual observed EVQ fraction. The derived overall detection fraction in the latter case is $\sim 0.4$, still consistent with what we found in the simple model approach\footnote{The detection fraction in the $|\Delta g|>0.8$ case is slightly higher than that in the $|\Delta g|>1$ case because the timescales of the extreme variability are on average shorter in the former case, and thus we suffer less from the selection incompleteness due to the limited observing baseline.}. 

\section{Conclusions}\label{sec:con}

We have performed a systematic search for extreme variability quasars (EVQs) using the combined SDSS and DES light curves over a time baseline of $\sim 15$ years. We found that there are $\sim 10\%$ ($2\%$) quasars show maximum $g$-band variability $>1$ ($1.5$) mag over this period, but this fraction is a strong function of luminosity. The intrinsic fraction of EVQs over this period, however, can be substantially higher ($\sim 30-50\%$) after correcting for selection incompleteness. 

These EVQs have slightly lower luminosities than the parent sample of the search. However, when compared to a control sample matched in luminosity and redshift, these EVQs display significant differences in their spectral line properties and variability properties. In particular, the narrow \OIII\ lines and broad \MgII\ and \CIV\ lines of the EVQs have larger EWs, and the EVQs are more variable on all timescales, compared to the control sample. Collectively these findings lead to the conclusion that EVQs have lower Eddington ratios than normal quasars matched in luminosity and redshift. 

Despite the difference in Eddington ratio (and consequently emission line properties), we do not find evidence that EVQs are a distinct population of quasars. There are continuous trends in the maximum variability as functions of quasar properties, suggesting that Eddington ratio is the main driver for a quasar to display extreme variability over multi-year timescales. 

We provide the list of EVQs identified from SDSS$+$DES. A subset of these objects are currently in the low-luminosity state. These objects are good candidates for Type 1 -- Type 2 CLQs, where the broad-line flux should drop substantially in the dim state. In addition, these recently dimmed quasars are good targets to observe their host galaxies since the nuclear emission is greatly reduced, and to study the correlations between quasar BH mass and host properties. 

In future work, we plan to incorporate additional photometric data in the light curves of SDSS$+$DES quasars and to recover more EVQs. In addition, we plan to study the multi-wavelength properties of EVQs in more details by taking advantage of the ample multi-wavelength data (such as X-ray data) in Stripe 82. Finally, we will extend our search to galaxies that recently turned on as quasars with multi-year photometric light curves. 

\acknowledgements

We thank Yan-Fei Jiang, Charles Gammie, and Aaron Barth for useful comments on the manuscript. YS acknowledges support from an Alfred P. Sloan Research Fellowship and NSF grant AST-1715579. 


Funding for the DES Projects has been provided by the U.S. Department of Energy, the U.S. National Science Foundation, the Ministry of Science and Education of Spain, 
the Science and Technology Facilities Council of the United Kingdom, the Higher Education Funding Council for England, the National Center for Supercomputing 
Applications at the University of Illinois at Urbana-Champaign, the Kavli Institute of Cosmological Physics at the University of Chicago, 
the Center for Cosmology and Astro-Particle Physics at the Ohio State University,
the Mitchell Institute for Fundamental Physics and Astronomy at Texas A\&M University, Financiadora de Estudos e Projetos, 
Funda{\c c}{\~a}o Carlos Chagas Filho de Amparo {\`a} Pesquisa do Estado do Rio de Janeiro, Conselho Nacional de Desenvolvimento Cient{\'i}fico e Tecnol{\'o}gico and 
the Minist{\'e}rio da Ci{\^e}ncia, Tecnologia e Inova{\c c}{\~a}o, the Deutsche Forschungsgemeinschaft and the Collaborating Institutions in the Dark Energy Survey.

The Collaborating Institutions are Argonne National Laboratory, the University of California at Santa Cruz, the University of Cambridge, Centro de Investigaciones Energ{\'e}ticas, 
Medioambientales y Tecnol{\'o}gicas-Madrid, the University of Chicago, University College London, the DES-Brazil Consortium, the University of Edinburgh, 
the Eidgen{\"o}ssische Technische Hochschule (ETH) Z{\"u}rich, 
Fermi National Accelerator Laboratory, the University of Illinois at Urbana-Champaign, the Institut de Ci{\`e}ncies de l'Espai (IEEC/CSIC), 
the Institut de F{\'i}sica d'Altes Energies, Lawrence Berkeley National Laboratory, the Ludwig-Maximilians Universit{\"a}t M{\"u}nchen and the associated Excellence Cluster Universe, 
the University of Michigan, the National Optical Astronomy Observatory, the University of Nottingham, The Ohio State University, the University of Pennsylvania, the University of Portsmouth, 
SLAC National Accelerator Laboratory, Stanford University, the University of Sussex, Texas A\&M University, and the OzDES Membership Consortium.

The DES data management system is supported by the National Science Foundation under Grant Number AST-1138766.
The DES participants from Spanish institutions are partially supported by MINECO under grants AYA2012-39559, ESP2013-48274, FPA2013-47986, and Centro de Excelencia Severo Ochoa SEV-2012-0234.
Research leading to these results has received funding from the European Research Council under the European Union's Seventh Framework Programme FP7/2007-2013) including ERC grant agreements 240672, 291329, and 306478.

Funding for the SDSS and SDSS-II has been provided by the Alfred P. Sloan Foundation, the Participating Institutions, the National Science Foundation, the U.S. Department of Energy, the National Aeronautics and Space Administration, the Japanese Monbukagakusho, the Max Planck Society, and the Higher Education Funding Council for England. The SDSS Web site is http://www.sdss.org/.

The SDSS is managed by the Astrophysical Research Consortium for the Participating Institutions. The Participating Institutions are the American Museum of Natural History, Astrophysical Institute Potsdam, University of Basel, University of Cambridge, Case Western Reserve University, University of Chicago, Drexel University, Fermilab, the Institute for Advanced Study, the Japan Participation Group, Johns Hopkins University, the Joint Institute for Nuclear Astrophysics, the Kavli Institute for Particle Astrophysics and Cosmology, the Korean Scientist Group, the Chinese Academy of Sciences (LAMOST), Los Alamos National Laboratory, the Max-Planck-Institute for Astronomy (MPIA), the Max-Planck-Institute for Astrophysics (MPA), New Mexico State University, Ohio State University, University of Pittsburgh, University of Portsmouth, Princeton University, the United States Naval Observatory, and the University of Washington.

Facility: DES, Sloan


\end{document}